\documentclass[12pt]{article}
\usepackage{latexsym}
\usepackage{amsmath}
\usepackage{amssymb}
\usepackage{relsize}
\usepackage{geometry}
\usepackage{tensor}
\usepackage{physics} 
\usepackage{csquotes} 

\usepackage{epsfig}
\usepackage{graphicx}       	
\graphicspath{{./img/}{./img/feynman/}{./img/numerical/}} 
\usepackage[font=small,labelfont=bf]{caption}
\usepackage{subcaption}

\def\beqn{\begin{eqnarray}}
\def\eeqn{\end{eqnarray}}

\def\beq{\begin{equation}}
\def\eeq{\end{equation}}
\def\ba{\beq\new\begin{array}{c}}
\def\ea{\end{array}\eeq}

\def\Tr{{\rm Tr}}
\newcommand{\gsim}{\lower.7ex\hbox{$
\;\stackrel{\textstyle>}{\sim}\;$}}
\newcommand{\lsim}{\lower.7ex\hbox{$
\;\stackrel{\textstyle<}{\sim}\;$}}

\newcommand{\ntwo}{${\mathcal N}=2$ }

\newcommand{\ntwot}{${\mathcal N}= \left(2,2\right) $ }
\newcommand{\ntwoo}{${\mathcal N}= \left(0,2\right) $ }
\newcommand{\none}{${\mathcal N}=1$ }

\newcommand{\pt}{\partial}

\newcommand{\bren}{{\beta_\text{ren}}}

\numberwithin{equation}{section}

\newcommand{\p}{\partial}
\newcommand{\wt}{\widetilde}
\newcommand{\ov}{\overline}

\def\slashed#1{\setbox0=\hbox{$#1$}             
   \dimen0=\wd0                                 
   \setbox1=\hbox{/} \dimen1=\wd1               
   \ifdim\dimen0>\dimen1                        
      \rlap{\hbox to \dimen0{\hfil/\hfil}}      
      #1                                        
   \else                                        
      \rlap{\hbox to \dimen1{\hfil$#1$\hfil}}   
      /                                         
   \fi}                                        %

\newcommand{\nbar}{\ov{n}}

\newcommand{\CP}{CP($N-1$)~}

\usepackage{mathtools}
\usepackage{hyperref}
\usepackage{xcolor}
\usepackage{tikz-cd}

\begin{document}

\hypersetup{%
	linkbordercolor=blue,
}

%
%

\begin{titlepage}

\begin{center}
	\Large{{\bf Dynamics of non-Abelian strings in the theory interpolating from \ntwo to  \none supersymmetric QCD
	}}
	
	\vspace{5mm}
	
	\vspace{5mm}
	
{\large \bf A.~Gorsky$^{\,a,b}$, \bf E.~Ievlev$^{\,c,d}$ and  A.~Yung$^{\,c,e}$}
\end{center}
\begin{center}
$^{a}${\it Institute for Information Transmission Problems of RAS, Russia}\\
	
	$^{b}${\it Moscow Institute of Physics and Technology, Dolgoprudnyi, Russia}\\

	$^{c}${\it National Research Center ``Kurchatov Institute'',
	Petersburg Nuclear Physics Institute, Gatchina, St. Petersburg
	188300, Russia}\\
	$^{d}${\it  St. Petersburg State University,
	 Universitetskaya nab., St. Petersburg 199034, Russia}\\
	 $^e${\it  William I. Fine Theoretical Physics Institute,
	University of Minnesota,
Minneapolis, MN 55455}\\
\end{center}

\vspace{1cm}

\begin{abstract}

We study the dynamics of non-Abelian vortex strings supported in  \ntwo  supersymmetric QCD with the U$(N)$ gauge group and $N_f=N$ quark flavors deformed
by the mass $\mu$ of the adjoint matter. In the limit of large $\mu$ the bulk four-dimensional theory flows to \none supersymmetric QCD. The dynamics of orientational zero modes of the non-Abelian string is
described by the world sheet CP$(N-1)$ model. At $\mu=0$ this model has \ntwot supersymmetry while at large 
$\mu$ it flows to the non-supersymmetric CP$(N-1)$ model. We solve the world sheet model in the large $N$ 
approximation and find a rich phase structure with respect to the deformation parameter $\mu$ and quark
mass differences. The phases include two strong coupling phases and two Higgs phases. In particular,
the Higgs phase at small $\mu$ supports \CP model kinks representing confined monopoles of the bulk QCD,
while in the  large-$\mu$ Higgs phase monopoles disappear.


\end{abstract}

\end{titlepage}

\newpage

\tableofcontents
\clearpage

%
%

\section*{Introduction} \label{introduction}
\addcontentsline{toc}{section}{Introduction}

Non-Abelian vortex strings were first found in \ntwo supersymmetric QCD (SQCD) with gauge group U$(N)$ and 
$N_f=N$ flavors of quark hypermultiplets \cite{HT1,ABEKY,SYmon,HT2}, see \cite{Trev,Jrev,SYrev,Trev2} for a reviews. When this theory is in the Higgs phase for scalar quarks (in the quark vacuum) non-Abelian strings are formed. They give rise to the confinement of  monopoles at weak coupling and to the so called ''instead-of-confinement'' phase for quarks at strong coupling, see \cite{SYdualrev} for a review. This picture gives a non-Abelian generalization of the Seiberg-Witten scenario of the Abelian quark confinement in the  monopole vacuum of \ntwo SQCD \cite{SW1,SW2}.

Besides translational zero modes typical for Abrikosov-Nielsen-Olesen (ANO) strings \cite{ANO} non-Abelian strings have orientational zero modes. Their internal dynamics is described by two-dimensional \ntwot supersymmetric  \CP model living on the string world sheet \cite{HT1,ABEKY,SYmon,HT2}. 

A lot of work has been done to generalize the construction of non-Abelian strings to QCD-like theories with less supersymmetry, in particular to \none SQCD \cite{SYnone,Edalati,SY02,YIevlevN=1} see \cite{SYrev} for a review.
One promising approach is to deform \ntwo SQCD by the mass $\mu$ of the adjoint matter ($\mu$-deformed SQCD) and study what happen
to non-Abelian strings upon this deformation. This deformation breaks \ntwo supersymmetry and in the limit of
$\mu\to\infty$ the bulk theory flows to \none SQCD. In Ref. \cite{YIevlevN=1} the world sheet theory living on the non-Abelian string in the $\mu$-deformed SQCD was found and it was shown that it flows to the non-supersymmetric \CP model in the limit of large $\mu$.

Since the bulk SQCD is in the Higgs phase for scalar quarks monopoles are confined by non-Abelian strings.
However, 
the monopoles cannot be attached to the string endpoints. In fact, in the U$(N)$ theories confined  
 monopoles 
are  junctions of two distinct elementary non-Abelian strings. From the point of view of 
\CP model living on the string world sheet confined monopoles are seen as kinks interpolating between
different vacua of \CP model \cite{Tong,SYmon,HT2} (see \cite{SYrev} 
for a review). 

In this paper we present a large $N$ solution of the world sheet theory for the non-Abelian string in the 
$\mu$-deformed SQCD. Large $N$ approximation was first used by Witten to solve both non-supersymmetric 
and \ntwot supersymmetric two-dimensional \CP models \cite{W79}. In particular, large-$N$ Witten's solution
shows that an auxiliary U(1) gauge field $A_{\mu}$ introduced to formulate \CP model becomes physical.
The \ntwot supersymmetric \CP model
has $N$ degenerate vacua as dictated by its Witten index. The order parameter which distinguishes between these vacua is the vacuum expectation value (VEV) of the scalar superpartner $\sigma$ of the gauge field $A_{\mu}$ \cite{W79}. 

In the non-supersymmetric
\CP model these vacua are split with splittings proportional to $1/N$ and become quasivacua. The theory has a single true vacuum \footnote{We assume below that the $\theta$-angle is zero.}. The order parameter which distinguish between these quasivacua is the value of the constant  field strength of the gauge field $A_{\mu}$ which  is massless in the non-supersymmetric case \cite{W79}, see also \cite{GSY05} and review \cite{SYrev}.

In this paper we use the large $N$ approximation to study  a  phase structure of the world sheet theory on the non-Abelian string in $\mu$-deformed SQCD with respect to the deformation parameter $\mu$ and quark
mass differences $\Delta m$. We find a rich phase structure which  includes two strong coupling phases and two Higgs phases. 

Strong coupling phases appear at small $\Delta m$. The first strong coupling phase appears at small $\mu$.
It is qualitatively similar to \ntwot supersymmetric phase at $\mu=0$. Although $N$ vacua are split
and become quasivacua the order parameter is still the VEV of the field $\sigma$. In the second 
strong coupling phase at large $\mu$ quasivacua are distinguished by the value of the constant electric field.
This phase is qualitatively similar to the non-supersymmetric \CP model.

At large $\Delta m$ we find two weakly coupled Higgs phases. At small $\mu$ $N$ vacua present in \ntwot case
split and become quasivacua. Still we have kinks interpolating between them.
As we increase $\mu$ above certain critical value, these lifted quasivacua disappear one by one, so we have a cascade of phase transitions. In the end we are left with a single vacuum and no kinks at all.

From the point of view of the bulk SQCD we interpret this as follows.
At large $\Delta m$ and small $\mu$ we have monopoles confined by non-Abelian strings while as we increase $\mu$
 monopoles disappear.

The paper is organized as follows. In Sec.~\ref{sec:revCP} we review non-supersymmetric and \ntwot supersymmetric \CP models
and their large-$N$ solutions. We also formulate the world sheet \CP model for non-Abelian string in $\mu$-deformed  SQCD. In Sec.~\ref{sec:Veff} we derive the  effective potential of this model in the large $N$ approximation.
Sec.~\ref{sec:strong} is devoted to the discussion of the two strong coupling phases at small $\Delta m$, while in
Sec.~\ref{sec:higgs} we study  Higgs phases at large $\Delta m$. In Sec.~\ref{branes}  we make  brief comments  on 
the brane picture of non-Abelian strings and 2d-4d  correspondence.
Sec.~\ref{Conclusions} contains description of the phase diagram of the world sheet model and  our conclusions.

%
%

\section{Review of \CP sigma models \label{sec:revCP}}

In this section we define basic \CP models that are of interest to us.
First, we will briefly review the non-supersymmetric and the \ntwot supersymmetric models, which were considered before, see for example \cite{W79,W93,Gorsky:2005ac,BSY3}. After that, we will introduce the model that we will be working with, namely the $\mu$-deformed \CP model which is an effective  theory living on the world sheet
of non-Abelian string in $\mu$-deformed SQCD \cite{YIevlevN=1}. 

\subsection{Non-supersymmetric model}

Throughout this paper we will be working with the gauge formulation \cite{W79} of the \CP models. In this formalism, the model is formulated via  $N$ complex scalar fields $n^i$, $i = 1, \ldots, N$ interacting with auxiliary U(1) gauge field $A_{\mu}$. The Lagrangian is written as
\begin{equation}
\mathcal L =
	\left|\nabla_\mu n^i\right|^2  
	+ i\,D\left(\bar{n}_i n^i -2\beta_0 \right)+ \sum_i\left|\sqrt 2\sigma-m_i\right|^2\, |n^i|^2,
\label{cpn_lagr_simplest}	
\end{equation}
where $\nabla_\mu = \partial_\mu -i\,  A_\mu$. Fields $\sigma$ and $D$ come without kinetic energy and are also auxiliary. They can be eliminated via their equations of motion. In particular integrating out $D$ imposes 
the constraint
\begin{equation}
	\nbar_i \, n^i = 2\beta_0, 
\label{cp_constraint_n}	
\end{equation}
which together with gauge invariance reduces the number of real degrees of freedom of the $n^i$ field down to
$2(N-1)$.

This is the non-supersymmetric  version of the \CP model, and it arises as a world sheet theory on the non-Abelian string in a non-supersymmetric QCD-like theory, see \cite{GSY05} and review \cite{SYrev}. The mass parameters $m_i$ are equal to quark masses in the four-dimensional  theory. 

Throughout this paper we will consider the masses placed uniformly on a circle,
\begin{equation}
	m_k = m - \Delta m \, \exp(\frac{2 \pi i \, k}{N}), \quad k=0,\, \ldots,\, N-1 \,.
\label{masses_ZN}	
\end{equation}
Here $m \in \mathbb{R}$ is the average mass, and $\Delta m > 0$ is effectively the mass scale of the  model.
Note that by  a shift of $\sigma$ one can always add a  constant to all 
 $m_i$. In particular one can get rid of the average mass $m$.

The bare coupling constant $\beta_0$ in quantum theory becomes a running coupling $\beta$. It is asymptotically free and defines the scale $\Lambda_{CP}$ via 
\begin{equation}
	\Lambda_{CP}^2 = M_\text{uv}^2 \, \exp(- \frac{8 \pi \beta_0}{N}),
\label{2d_4d_coupling_nosusy}	
\end{equation}
where $M_\text{uv}$ is the ultra-violet (UV) cutoff.

Let us review phases of this theory. It is known 
that in the case of vanishing masses $\Delta m = 0$ this non-supersymmetric \CP model is at strong coupling with vanishing VEV $\langle n^i \rangle = 0$. It can be solved by means of the $1 / N$ expansion \cite{W79}. It turns out that at the quantum level spontaneous breaking of the global
SU(N) (flavor) symmetry present at the classical level disappears. There are no massless Goldstone bosons in the physical
spectrum. The $n^i$ fields acquire mass of the order of $\Lambda_{CP}$. 

Moreover,  composite degree of freedom -- the would-be auxiliary photon $A_\mu$ acquires a kinetic term at the one-loop level and becomes dynamical. The presence of massless photon ensures long range
forces in the non-supersymmetric \CP model. The Coulomb potential is linear
in two dimensions, namely 
\begin{equation}
	V(r) \sim \frac{\Lambda_{CP}^2}{N} \, r \,,
\end{equation}
where $r$ is the separation between the charges.
This leads to the Coulomb/confinement phase \cite{W79}. Electric charges
are confined. The lightest electric charges are the $n^i$ quanta
which become kinks at strong coupling \cite{W79}. Confinement of kinks
means that they are not present in the physical spectrum of the theory in isolation.
They form bound states, kink-antikink \textquote{mesons}. 

\begin{figure}[h]
	\centering
	\includegraphics[width=0.5\linewidth]{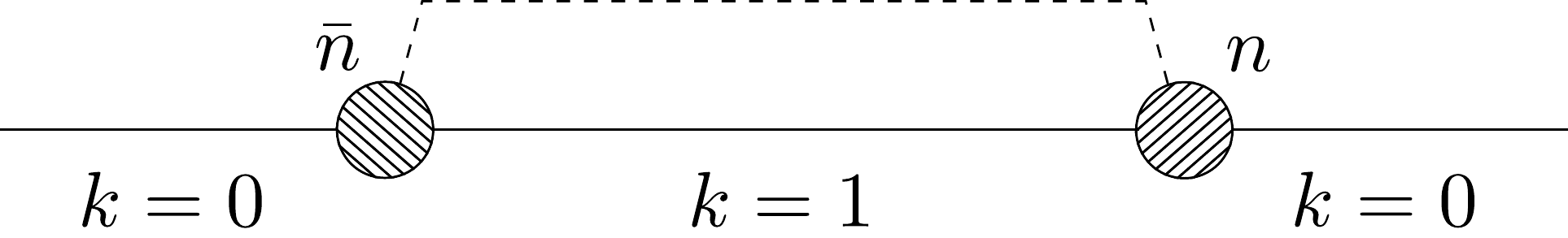}
	\caption{
	Linear confinement of the $n$-$\bar{n}$ pair.
	The solid straight line represents the ground state ($k=0$ vacuum).
	The dashed line shows
	the vacuum energy density in the first quasivacuum.}
\label{fig:conf}
\end{figure}

Masses of kinks are of order of $\Lambda_{CP}$ while the confining potential is weak, proportional to $1/N$.
Therefore kink and antikink in the ''meson'' are well separated forming a quasivacuum inside the ''meson''. Thus,
 beside the single ground state, there is a family of quasivacua with energy splittings of order $\sim \Lambda_{CP}^2 / N$. The order parameter which distinguish different quasivacua is the value of the constant
electric field or topological density
\beq
Q=\frac{i}{2\pi}\,\varepsilon_{\mu\nu}\,\pt^{\mu}A^{\nu} =\frac1{8\pi\beta} \,\varepsilon_{\mu\nu}
\,\pt^{\mu}\bar{n}_i\pt^{\nu}n^i
\label{topdensity}
\eeq
The picture of
confinement of $n$'s is shown on Fig.~\ref{fig:conf}.


The kinks interpolate between the adjacent vacua. They are confined
monopoles of the bulk theory. Since the excited string tensions are larger than the
tension of the lightest one, these monopoles, besides four-dimensional confinement,
are confined also in the two-dimensional sense: a monopole is necessarily attached to
an antimonopole on the string to form a meson-like configuration \cite{MMY,GSY05}
 
On the other hand, at large mass scales $\Delta m \gg \Lambda_{CP}$ the coupling constant is small, frozen at the scale $\Delta m$, and semiclassical calculations are applicable. The field $n^i$ develops a non-zero VEV, and there is no massless photon and no long-range interactions. That is why this phase is usually called \textquote{Higgs phase} as opposed to the Coulomb/confinement strong coupling phase. More exactly \CP model
in this phase gives a low energy description of a Higgs phase below the scale of the photon mass. Essentially
this weakly coupling  Higgs phase is similar to the ''classical phase'' described by the classical Lagrangian 
\eqref{cpn_lagr_simplest}.

It was shown that at intermediate mass scales $\Delta m \sim \Lambda_{CP}$ there is a phase transition between the Higgs and Coulomb phases, see \cite{Gorsky:2005ac, GSY05, Ferrari, Ferrari2}.

\subsection{\ntwot model}

Supersymmetric generalization of the above model \cite{W79,W93} has additional fermionic field $\xi^i$, $i = 1, \ldots, N$, which are superpartners of the $n_i$ fields. 
The Euclidean version of the full \ntwot Lagrangian is 
\begin{equation}
\begin{aligned}
\mathcal L =&
	\frac{1}{e_0^2}\left(\frac{1}{4} F_{\mu\nu}^2 +\left|\pt_\mu\sigma\right|^2 + \frac{1}{2}D^2
	+\bar\lambda \, i\bar{\sigma}^\mu\pt_\mu\,\lambda
	\right) + i\,D\left(\bar{n}_i n^i -2\beta_0
	\right)
	\\
	&+
	\left|\nabla_\mu n^i\right|^2+ \bar{\xi}_i\, i\bar{\sigma}^\mu\nabla_\mu\,\xi^i
	+ 2\sum_i\left|\sigma-\frac{m_i}{\sqrt 2}\right|^2\, |n^i|^2
	\\
	&+
	i\sqrt{2}\,\sum_i \left( \sigma -\frac{m_i}{\sqrt 2}\right)\bar\xi_{Ri}\, \xi^i_L 
	- i\sqrt{2}\,\bar{n}_i \left(\lambda_R\xi^i_L - \lambda_L\xi^i_R \right)
	\\
	&+
	i\sqrt{2}\,\sum_i \left( \bar\sigma -\frac{\bar{m}_i}{\sqrt 2}\right)\bar\xi_{Li}\, \xi^i_R 
	- i\sqrt{2}\,{n}^i \left(\bar\lambda_L\bar\xi_{Ri} - \bar\lambda_R\bar\xi_{Li} \right),
\end{aligned}	
\label{lagrangian_N=2}
\end{equation}
where  $m_i$ are twisted masses and  the limit $e_0^2\to\infty$ is implied. Moreover, $\bar\sigma^\mu = \{1,\,i\sigma_3\}$ . Fermions $\xi_L, \xi_R$ are respectively left and right components of the $\xi$ field. Here again one can add a uniform constant to all the $m_i$ by shifting the $\sigma$ field.

The gauge field $A_\mu$, complex scalar superpartner $\sigma$, real scalar $D$ 
and a two-component complex fermion  $\lambda$ form a vector  auxiliary supermultiplet. 
In particular, integrating over $D$ and fermion $\lambda$ give
the  constraints 
\begin{equation}
	\nbar_i \, n^i = 2\beta_0,
\end{equation}
\begin{equation}
	\bar{n}_i\,\xi^i =0\,,\qquad \bar\xi_i\,n^i = 0\,
\end{equation}
in the limit $e_0\to\infty$.

This model was derived as a world sheet theory on the non-Abelian string in \ntwo SQCD. The $n_i$ fields parametrize  the orientational moduli of the non-Abelian string \cite{HT1,ABEKY,SYmon,HT2} 
The mass parameters $m_i$ are in fact masses of the bulk quark fields. The bare coupling constant $\beta_0$ is related to the bulk gauge coupling constant $g^2$ normalized at the scale of the bulk gauge boson mass $m_G \sim g\sqrt{\mu m}$ via (see e.g.~\cite{SYrev})
\begin{equation}
	2\beta_0 = \frac{4\pi}{g^2(m_G)} = \frac{N}{2\pi}\ln{\frac{m_G}{\Lambda_{CP}}}\,,
\label{2d_4d_coupling_susy}	
\end{equation}
In order to keep the bulk theory at weak coupling we assume that $m_G \gg \Lambda_{CP}$. 

Witten solved this model in the large $N$ approximation in the zero mass case  \cite{W79}.
Large-$N$ solution of this model at non-zero masses shows two different regimes at weak and  strong coupling \cite{Bolokhov:2010hv}. At small mass scales $\Delta m < \Lambda_{CP}$ the theory is in the strong coupling phase with zero VEV $\langle n^i \rangle = 0$ and with a dynamical photon (Witten's phase). However the photon  now is massive due to the presence of the chiral anomaly. There are no long-range forces and no confinement of kinks. 

In both strong and weak coupling regimes the 
 theory has $N$ degenerate vacuum states as dictated by its Witten index. They are  labeled by the VEV of 
$\sigma$ \cite{Bolokhov:2010hv}. At $\Delta m <\Lambda_{CP}$ we have 
\begin{equation}
	\sqrt{2}\sigma ~=~ \exp\left( \frac{2\pi\,i\, k}{N}
	\right)\times \Lambda_{CP}
	 \qquad k=0, ..., N-1
	\label{22sigmaapp}
\end{equation}
This result can be understood as follows.
The chiral anomaly breaks U(1) $R$-symmetry present at zero masses down to $Z_{2N}$ which is then broken spontaneously down to $Z_2$ by VEV of the $\sigma$ field (which has $R$ charge equal to two). 
In particular, from the large-$N$ solution it follows that VEV of  $\sqrt{2}|\sigma| = \Lambda_{CP}$.
Then 
$Z_{2N}$
symmetry ensures presence of $N$ vacua as shown in  Eq.~\eqref{22sigmaapp}. 

At large masses located on a circle
(see \eqref{masses_ZN}) the $Z_{2N}$ symmetry is still unbroken. This  leads to to the similar structure of the $\sigma$  VEVs at $\Delta m > \Lambda_{CP}$, namely
\begin{equation}
	\sqrt{2}\sigma ~=~ \exp\left( \frac{2\pi\,i\, k}{N}
	\right)\times \Delta m,
	 \qquad k=0, ..., N-1
	\label{sigmavevLargem}
\end{equation}

The above  formulas show a phase transition at $\Delta m = \Lambda_{CP}$. 
As follows from the large-$N$ solution the model above this point  is in the Higgs phase with a nonzero VEV for say, zero component of $n$, $\langle n^0 \rangle \neq 0$.  In both phases there is no confinement, in contrast  to the non-supersymmetric case.

In fact the above phase transition is a consequence of the large-$N$ approximation. At finite $N$ the transition 
between two regimes is smooth. This follows from the exact effective superpotential known for \ntwot \CP
model \cite{W93}.

\subsection{$\mu$-Deformed \CP model \label{sec:intro_deformed}}

Now let us pass on to the case of interest, namely, the $\mu$-deformed \CP model. 
This model appears as a world sheet theory on a non-Abelian string in \ntwo SQCD deformed by the adjoint field mass $\mu$. It was derived in \cite{YIevlevN=1} in two cases, for small and large values of the deformation $\mu$.
Here and throughout this paper we will take the mass parameters to lie on the circle \eqref{masses_ZN}, and we also assume that the deformation parameter is real and positive, $\mu > 0$. 
%

The first effect derived  in \cite{YIevlevN=1} is  that  $n_i$ fields entering the \ntwot \CP model \eqref{lagrangian_N=2} develop an additional potential  upon $\mu$ deformation which depends on mass differences. This potential in the small $\mu$ limit was first found in \cite{SYfstr}.
The second effect is that superorientational modes of the non-Abelian string are lifted. In other words the
two-dimensional fermions $\xi^i$ (fermionic superpartners of $n^{i}$)  were massless in the supersymmetric version of the model at $\mu=0$. However, at  small $\mu$ they  acquire a mass $\lambda(\mu) \sim \mu$ \cite{YIevlevN=1}. At large deformations they become heavy and decouple.

In order to capture these features, we write the following Lagrangian for the deformed \CP model:
\begin{equation}
\begin{aligned}
\mathcal L &= 
\left|\nabla_\mu n^i\right|^2+ \bar{\xi}_i\, i\bar{\sigma}^\mu\nabla_\mu\,\xi^i 
+ i\,D\left(\bar{n}_i n^i -2\beta \right)
\\[3mm]
&+ \sum_i\left|\sqrt 2\sigma-m_i\right|^2\, |n^i|^2
+ \upsilon (\mu) \sum_i \Re\Delta m_{i0} |n^i|^2
\\[3mm]
&+
i\,\sum_i \left( \sqrt{2}\sigma -m_i - \lambda (\mu) \right)\bar\xi_{Ri}\, \xi^i_L 
- i\sqrt{2}\,\bar{n}_i \left(\lambda_R\xi^i_L - \lambda_L\xi^i_R \right)
\\
&+
i\,\sum_i \left( \sqrt{2}\bar\sigma - \bar{m}_i - \ov{\lambda (\mu)}  \right)\bar\xi_{Li}\, \xi^i_R 
- i\sqrt{2}\,{n}^i \left(\bar\lambda_L\bar\xi_{Ri} - \bar\lambda_R\bar\xi_{Li} \right),
\end{aligned}
\label{lagrangian_init}
\end{equation}
where $\Delta m_{i0} = m_i - m_0$,  $m_i$ are  quark masses $i=1,...N$, and $m_0$ is the mass with the smallest real part.

The coefficient functions $\upsilon (\mu)$ and $\lambda (\mu)$ were derived in \cite{YIevlevN=1} at the classical level for small and large values of $\mu$:
\begin{equation}
\upsilon (\mu) = 
	\begin{cases}
		\frac{4\pi\mu}{2\beta} 											\,, \quad &\mu \to 0 ,\\
		\frac{1}{2\beta} \frac{8\pi \mu}{\ln \frac{g^2\mu}{m}} 	\,, \quad &\mu \to \infty
	\end{cases}
\label{upsilon_old}	
\end{equation}
\begin{equation}
\lambda (\mu) =
	\begin{cases}
		\lambda_0 \frac{\mu}{2\beta} \,, \quad &\mu \to 0 ,\\
		\text{const  } g\sqrt{\mu m} \sim m_G   \,, \quad &\mu \to \infty
	\end{cases}
\label{lambda_old}	
\end{equation}
%
%
Here $g^2$ is the four-dimensional  bulk coupling constant.
The numerical value for $\lambda_0$ is  $\lambda_0 \approx 3.7$ \cite{YIevlevN=1}. Note that although we can get rid 
of the explicit dependence on the average quark mass $m$ in \eqref{lagrangian_init} by shift of $\sigma$ the above formulas show that  it enters indirectly through definitions of parameters of $\mu$-deformed \CP model \eqref{lagrangian_init} in terms of parameters of the bulk SQCD.

This model interpolates between the supersymmetric and the non-supersymmetric models briefly described above. In the limit $\mu \to 0$ supersymmetry is restored to \ntwot, and we obtain \eqref{lagrangian_N=2}. At large deformations the fermions can be integrated out, and the theory flows to the bosonic model \eqref{cpn_lagr_simplest}. 

Our main tool of investigating this model in the quantum level will be the $1 / N$ expansion. In order to have a smooth large $N$ limit, our parameters should scale as
\begin{equation}
\begin{aligned}
	g^2   \sim 1/N , \quad
	\beta \sim N , \quad
	\mu   \sim N , \\
	m \sim 1, \quad
	\upsilon(\mu) \sim 1 , \quad
	\lambda(\mu)  \sim 1
\end{aligned}
\end{equation}

Below in this paper we will use  three independent  physical parameters to describe  our four-dimensional bulk  model. The first one is the bulk gauge boson mass 
\beq
m_G^2=2 g^2 \mu m, 
\label{Wmass}
\eeq
which plays a role of the physical UV cutoff in the world sheet \CP model on the non-Abelian string, see \cite{SYrev}. The second one is 
the quark mass differences $( m_i-m_j)$ and the third parameter is the physical mass of the adjoint matter
\beq
m_\text{adj} = g^2\mu = \frac{\mu}{\frac{N}{8\pi^2}\, \ln\frac{m_G}{\Lambda_\text{4d}}} \equiv \wt{\mu} \,.
\label{tildem}
\eeq
which will be  our actual deformation parameter. All three parameters scales as $N^0$ in the large $N$ limit.
Here $\Lambda_{4d}$ is the scale of \ntwo bulk SQCD.

Thus, in fact, the average quark mass $m$ is not an independent parameter. It can be written  as 
\begin{equation}
	m  = \frac{m_G^2}{2\wt{\mu}} \,.
\end{equation}

At the scale of the gauge boson mass \eqref{Wmass} the world sheet coupling constant  for small $\mu$  is given by \cite{ABEKY, SYmon}, cf. \eqref{2d_4d_coupling_susy}
\begin{equation}
	2\beta = \frac{4\pi}{g^2}  = \frac{N}{2\pi}\, \ln\, \frac{m_G}{\Lambda_{4d}} . 
\end{equation}

For large $\mu$ the world sheet coupling normalized at the scale $m_G$ becomes \cite{YIevlevN=1}
\begin{equation}
	2\beta = \text{const  } \frac{\mu}{m}\,\frac{1}{\ln^2 \frac{g^2\mu}{m}}.  
\end{equation}
Expressed in terms of the invariant parameters it reads
\begin{equation}
	2\beta = \text{const  }\frac{ N}{\pi} \, \frac{\wt{\mu}^2}{m_G^2}  \frac{\ln \frac{m_G}{\Lambda_\text{4d}^{{\mathcal N}=1}}}{\ln^2 \frac{2 \wt{\mu}}{m_G}},
	\label{beta_classical_largemu}
\end{equation}
where we take into account that at large $\wt{\mu}$ our bulk theory flows to  \none SQCD  with the scale
$(\Lambda_\text{4d}^{{\mathcal N}=1})^{2N}=\wt{\mu}^N\Lambda_\text{4d}^N$.
 
In terms of the independent parameters the coefficient functions $\upsilon$ and $\lambda$ become
\begin{equation}
\upsilon (\wt{\mu}) = 
	\begin{cases}
		\wt{\mu} 											\,, \quad &\wt{\mu} \to 0 ,\\
		\frac{m_G^2}{\wt{\mu}} \, \ln \frac{2 \wt{\mu}}{m_G} 		 	\,, \quad &\mu \to \infty
	\end{cases}
\label{upsilon}	
\end{equation}
\begin{equation}
\lambda (\wt{\mu}) =
	\begin{cases}
		\wt{\lambda}_0 \wt{\mu} \,, \quad &\wt{\mu} \to 0 ,\\
		m_G   \,, \quad &\wt{\mu} \to \infty
	\end{cases}
\label{lambda}	
\end{equation}
where $\wt{\lambda}_0 = \lambda_0 / 4\pi \approx 0.3$.

As we already mentioned the value of the bulk gauge boson mass $m_G$ plays a role of the UV cutoff of our world sheet theory.
Below $m_G$ our model is asymptotically free (cf. \eqref{2d_4d_coupling_nosusy}) with 
\beq
2\beta(E) = \frac{N}{2\pi}\, \ln\, \frac{E}{\Lambda_{2d}} 
\label{beta_run}
\eeq
at the scale $E$.
This fixes the scale $\Lambda_{2d}$ in terms of the parameters of the bulk theory. At small $\wt{\mu}$ we have 
\beq
\Lambda_{2d}(\wt{\mu}\to 0) = \Lambda_{4d}, 
\label{Lambda2D=4D}
\eeq
while at large $\wt{\mu}$ 
\begin{equation}
	\Lambda_{2d} = \Lambda_\text{4d}^{{\mathcal N}=1} 
	\exp(- \text{const}\,\frac{ \wt{\mu}^2}{m_G^2} \cdot \frac{1}{\ln \frac{2 \wt{\mu}}{m_G}})
\label{Lam_2d}	
\end{equation}


 Note that at  $\wt{\mu} \to \infty$ the scale \eqref{Lam_2d} of our model becomes exponentially
small and the model enters the strong coupling regime only at extremely small energies. We will see below that phase transitions with respect to $\wt{\mu}$ appear at rather small values of $\wt{\mu}$ where the scale $\Lambda_{2d}$ is close to its supersymmetric value $\Lambda_{4d}$. Since the fermion decoupling occurs at very large $\wt{\mu} \gg m_G$, we can use small $\wt{\mu}$ approximation formulas \eqref{upsilon} and \eqref{lambda} while investigating the phase transition.

In the following sections we are going to investigate different phases and vacuum structure of the world sheet  theory. There are two parameters that we can vary -- the SUSY breaking parameter $\wt{\mu}$ and the mass scale $\Delta m$. As we already mentioned  our model \eqref{lagrangian_init} exhibits a rich phase structure  in the $(\Delta m, \ \wt{\mu})$ plane.

%
%

\section{One loop effective potential \label{sec:Veff}}

In this section we proceed with solving the theory \eqref{lagrangian_init} via the $1 / N$ expansion. As we already mentioned the \ntwot model as well as the non supersymmetric \CP model (without mass parameters) were solved by Witten \cite{W79}. This method was also generalized for the case of heterotic \ntwoo model \cite{SYhetN} and for the twisted mass case \cite{Gorsky:2005ac,Bolokhov:2010hv}. Our derivation will closely follow these papers.

\subsection{Derivation of the effective potential}

We want to start by deriving the one-loop effective potential. Our action \eqref{lagrangian_init} is well suited for that since it is quadratic with respect to the dynamical fields $n_i$ and $\xi_i$. However, we do not to integrate over all of them a priori due to  the  following reason.

As was stated in the previous section, our model \eqref{lagrangian_init} is, in a sense, an intermediate case between the \ntwot and the non-supersymmetric \CP models, which were studied before. Therefore we can use the insights derived from these models in order to better understand our case. First of all, we expect that our theory has at least two phases, the strong and weak coupling. The order parameter distinguishing between these two phases is the expectation value of the $n_i$ fields. At weak coupling (so-called Higgs phase \cite{Gorsky:2005ac}) one of the $n_i$ develops a VEV, $\langle n_{i_0} \rangle = 2\beta$. In the strong coupling regime (so-called Coulomb phase), VEVs of all the $n_i$ field vanish.

So, we will use the following strategy. We integrate over  $N-1$ fields $n^{i}$ with $i \ne 0$ (and over the corresponding fermions $\xi_i$). The resulting effective action is a functional of $n^0\equiv n$, $D$ and $\sigma$. To find the vacuum configuration, we will minimize the effective action with respect to $n$, $D$ and $\sigma$.

Note  that this functional also depends on $A_\mu$ and the fermions $\xi_{L,R}^0$, $\lambda_{L,R}$, but  the Lorenz invariance imply that these fields have zero VEVs. We also choose to allow  $n^0$ field to have non-zero VEV because the associated mass $m_0$ has  the minimal real part (see \eqref{masses_ZN} ) and as we will see later  $\langle n^0 \rangle \neq 0$ corresponds to the true vacuum in the Higgs phase rather then a quasivacuum.

Integrating out the $n^i$ and $\xi^i$ fields, we arrive at the following determinants:
\begin{equation}
	 \frac{
	\prod_{i=1}^{N-1} {\rm det}\, \left(-\pt_{k}^2 
	   + \bigl| \sqrt{2}\sigma - m_i - \lambda (\mu) \bigr|^2\right)}{
	\prod_{i=1}^{N-1}{\rm det}\, \left(-\pt_{k}^2 +iD + \upsilon (\mu)\Delta m_{i0} 
	   + \bigl| \sqrt{2}\sigma - m_i \bigr|^2\right)},
\label{det}	
\end{equation}
which gives for the effective potential:
\begin{equation}
\begin{aligned}
	V_\text{eff} &= \int d^2 x\, (iD + |\sqrt{2}\sigma - m_0|^2) |n|^2 - 2\beta \int d^2 x \, iD \\
					&+ \sum_{i=1}^{N-1} \Tr \ln \left(-\pt_{k}^2 +iD + \upsilon (\mu)\Delta m_{i0}  + \bigl| \sqrt{2}\sigma - m_i \bigr|^2\right) \\
					&- \sum_{i=1}^{N-1} \Tr \ln \left(-\pt_{k}^2 + \bigl| \sqrt{2}\sigma - m_i - \lambda (\mu) \bigr|^2\right)
\end{aligned}	
\end{equation}

The next step is to calculate the traces entering this expression. At $\wt{\mu} \to 0$, the supersymmetry is restored, and this expression is well defined. 
However at a non-vanishing deformation, this expression diverges quadratically, and a regularization needs to be performed. 
Below we proceed with the Pauli-Villars regularization (a similar procedure was carried out in \cite{NOVIKOV1984103}). We introduce regulator fields with masses $b_a$, $f_a$, $a=1, 2$, and write the regularized effective potential as
\begin{equation}
\begin{aligned}
	V_\text{eff} &= \int d^2 x\, (iD + |\sqrt{2}\sigma - m_0|^2) |n|^2 - 2\beta\, \int d^2 x \, iD \\
					&+ \sum_{i=1}^{N-1} \Tr \ln \left(-\pt_{k}^2 +iD + \upsilon (\mu)\Delta m_{i0}  + \bigl| \sqrt{2}\sigma - m_i \bigr|^2\right)	
					 + \sum_{a=1}^{2} \sum_{i=1}^{N-1} B_a \Tr \ln \left(-\pt_{k}^2 + b_a^2 \right) \\
					&- \sum_{i=1}^{N-1} \Tr \ln \left(-\pt_{k}^2 + \bigl| \sqrt{2}\sigma - m_i - \lambda (\mu) \bigr|^2\right)	
					 - \sum_{a=1}^{2} \sum_{i=1}^{N-1} F_a \Tr \ln \left(-\pt_{k}^2 + f_a^2\right)
\end{aligned}	
\end{equation}
where the coefficients satisfy
\begin{equation}
	\sum_{a=0}^{2} B_a = -1, \quad \sum_{a=0}^{2} B_a b_a^2 = - m_\text{bos}^2
\end{equation}
These equations imply
\begin{equation}
	B_1 = \frac{b_2^2 - m_\text{bos}^2}{b_1^2 - b_2^2}, \quad B_2 = - \frac{b_1^2 - m_\text{bos}^2}{b_1^2 - b_2^2}
\end{equation}
The regulator masses play the role of the UV cutoff. Similar relations hold for the fermionic regulator coefficients.

Moreover, we need to properly normalize our traces by subtracting the contributions in the trivial background, namely 
$\Tr \ln (-\pt_{k}^2)$ from the bosonic and  the fermionic traces. After this procedure we arrive at
\begin{equation}
\begin{aligned}
	V_\text{eff} &= \int d^2 x\, (iD + |\sqrt{2}\sigma - m_0|^2) |n|^2 - 2\beta\, \int d^2 x \, iD \\
					&- \frac{1}{4\pi} \sum_{i=1}^{N-1} \Bigg[ 
						\left( +iD + \upsilon (\mu)\Delta m_{i0}  + \bigl| \sqrt{2}\sigma - m_i \bigr|^2\right)
							\ln \left( +iD + \upsilon (\mu)\Delta m_{i0}  + \bigl| \sqrt{2}\sigma - m_i \bigr|^2\right) \\
					&- \left( +iD + \upsilon (\mu)\Delta m_{i0}  + \bigl| \sqrt{2}\sigma - m_i \bigr|^2\right)
							\frac{b_1^2 \ln b_1^2 - b_2^2 \ln b_2^2}{b_1^2 - b_2^2}
						\Bigg]	\\
					&+ \frac{1}{4\pi} \sum_{i=1}^{N-1} \Bigg[
						\bigl| \sqrt{2}\sigma - m_i - \lambda (\mu) \bigr|^2 \ln \bigl| \sqrt{2}\sigma - m_i - \lambda (\mu) \bigr|^2 \\
					&- \bigl| \sqrt{2}\sigma - m_i - \lambda (\mu) \bigr|^2 \frac{f_1^2 \ln f_1^1 - f_2^2 \ln f_2^2}{f_1^2 - f_2^2}
						\Bigg]
\end{aligned}	
\end{equation}

This is a quite complex expression.
In order to simplify it, let us take the limit \cite{NOVIKOV1984103}
\begin{equation}
	b_1^2 = x  M_\text{uv}^2 , \ b_2^2 = M_\text{uv}^2, \quad f_1^2 = x M_\text{uv}^2 , \ f_2^2 = M_\text{uv}^2, \quad x \to 1,
\end{equation}
where $M_\text{uv}$ is the UV cutoff.
%
%
Moreover, recall from the section \ref{sec:intro_deformed} that the bare coupling constant can be parametrized as
\begin{equation}
	2\beta (M_\text{uv}) ~~=~~\frac{N}{4\pi}\, \ln\, {\frac{M_\text{uv}^2}{\Lambda^2}}\,,
\end{equation}
Here, $\Lambda \equiv \Lambda_{2d}$ is the scale of our model. 
Then the density of the effective potential becomes
\begin{equation}
\begin{aligned}
	\mathcal V_\text{eff} &=  (iD + |\sqrt{2}\sigma - m_0|^2) |n|^2
					+ \frac{1}{4\pi} \sum_{i=1}^{N-1}  iD
							\left[ 1 - \ln\frac{iD + \upsilon (\mu) \Re\Delta m_{i0}  + \bigl| \sqrt{2}\sigma - m_i \bigr|^2}{\Lambda^2} \right]	\\
					&+ \frac{1}{4\pi} \sum_{i=1}^{N-1}  
						\left(\upsilon (\mu) \Re\Delta m_{i0}  + \bigl| \sqrt{2}\sigma - m_i \bigr|^2\right)
						\Bigg[ 1 - 
							\ln \frac{ iD + \upsilon (\mu) \Re\Delta m_{i0}  + \bigl| \sqrt{2}\sigma - m_i \bigr|^2}{M_\text{uv}^2} 
						\Bigg]\\
					&- \frac{1}{4\pi} \sum_{i=1}^{N-1} 
						\bigl| \sqrt{2}\sigma - m_i - \lambda (\mu) \bigr|^2 
						\Bigg[ 1 - 
							\ln \frac{\bigl| \sqrt{2}\sigma - m_i - \lambda (\mu) \bigr|^2}{M_\text{uv}^2}
						\Bigg]
\end{aligned}	
\label{v_eff}
\end{equation}

Note that our regularized effective potential depends on the UV cutoff scale $M_\text{uv}$. We cannot make a subtraction to get rid of it in the model at hand for the following reason. First, when we consider our
$\wt{\mu}$-deformed \CP model \eqref{lagrangian_init} as an effective world sheet theory  on the non-Abelian string the 
the UV cutoff has a clear physical meaning, namely
\beq
M_\text{uv} =m_G,
\label{M_UV}
\eeq
where $m_G$ is the mass of the bulk gauge boson. Moreover, the fermion mass $\lambda(\mu)$ in \eqref{v_eff} interpolates
from zero at $\wt{\mu}=0$ to $m_G=M_\text{uv}$ at $\wt{\mu}\to\infty$, see \eqref{lambda}. Thus $M_\text{uv}$ is in fact a physical parameter in our model and there is no need to get rid of it.

The renormalized coupling constant is 
\begin{equation}
	2\beta_\text{ren} = \frac{1}{4\pi}\sum_{i=1}^{N-1}\ln\frac{iD + \upsilon (\mu) \Re\Delta m_{i0}  + \bigl| \sqrt{2}\sigma - m_i \bigr|^2}{\Lambda^2}
\end{equation}

\subsection{Vacuum equations}

To find the vacuum configuration we minimize the effective potential \eqref{v_eff}. Varying with respect to $D$ we arrive at
\begin{equation}
	|n|^2 = 2\beta_\text{ren} = \frac{1}{4\pi}\sum_{i=1}^{N-1} \ln\frac{iD + \upsilon (\mu) \Re\Delta m_{i0}  + \bigl| \sqrt{2}\sigma - m_i \bigr|^2}{\Lambda^2}
\label{master1}
\end{equation}
Variation with respect to $\nbar$ yields the second equation:
\begin{equation}
	(iD + |\sqrt{2}\sigma - m_0|^2) n = 0
\label{master2}
\end{equation}
Finally, the third equation is obtained by minimizing over the $\sigma$ field,
\begin{equation}
\begin{aligned}
	- (\sqrt{2}\sigma - m_0) |n|^2 
			&+ \frac{1}{4\pi}\sum_{i=1}^{N-1} \left(\sqrt{2}\sigma - m_i\right) 
					\ln \frac{ iD + \upsilon (\mu) \Re\Delta m_{i0}  + \bigl| \sqrt{2}\sigma - m_i \bigr|^2}{m_G^2} \\
			&=  \frac{1}{4\pi}\sum_{i=1}^{N-1}\left(\sqrt{2}\sigma - m_i - \lambda (\mu)\right) 
					\ln \frac{\bigl| \sqrt{2}\sigma - m_i - \lambda (\mu) \bigr|^2}{m_G^2},
\end{aligned}
\label{master3}	
\end{equation}
where here and below we replaced $M_\text{uv}$ by the physical mass $m_G$.

These three equations comprise our {\em master set}. In addition, the vacuum configurations must satisfy the constraint
\begin{equation}
	\beta_\text{ren} \geqslant 0,
\label{beta_positive_condition}	
\end{equation}
which comes from $2\beta_\text{ren} = |n|^2 \geqslant 0$.

From \eqref{master1} and \eqref{master2} it immediately follows that either
\begin{equation}
	n = \beta_\text{ren} =  0
\label{strong_phase_condition}	
\end{equation}
or
\begin{equation}
	iD + |\sqrt{2}\sigma - m_0|^2 = 0 \, .
\label{higgs_phase_condition}	
\end{equation}
The first option corresponds to the strong coupling regime where the VEV of $n$ and the renormalized coupling constant vanish. The second option is realized in the Higgs regime, where the $n$ field develops a VEV. In the following sections we will study each of these regimes in detail.

%
%

\section{Strong coupling regime \label{sec:strong}}

In this section will begin investigation of our model in the strong coupling regime, which is defined by the condition \eqref{strong_phase_condition}. 
This phase occurs when the mass scale of the model $\Delta m \lesssim \Lambda$, see e.g. \cite{Gorsky:2005ac, Bolokhov:2010hv}. 
To start off we will first investigate a simple case $\Delta m = 0$. 
Behavior of our model is different at different values of the deformation parameter: at intermediate  $\wt{\mu}$ we will see a phase transition, while in the limit of large fermion mass $\lambda \to m_G$ we will confirm that the model \eqref{lagrangian_init} flows to non-supersymmetric \CP model \eqref{cpn_lagr_simplest} as expected.
Next, we will generalize our results to the case of distinct masses $m_i$.

\begin{figure}
    \centering
    \begin{subfigure}[t]{0.5\textwidth}
        \centering
        \includegraphics[width=\textwidth]{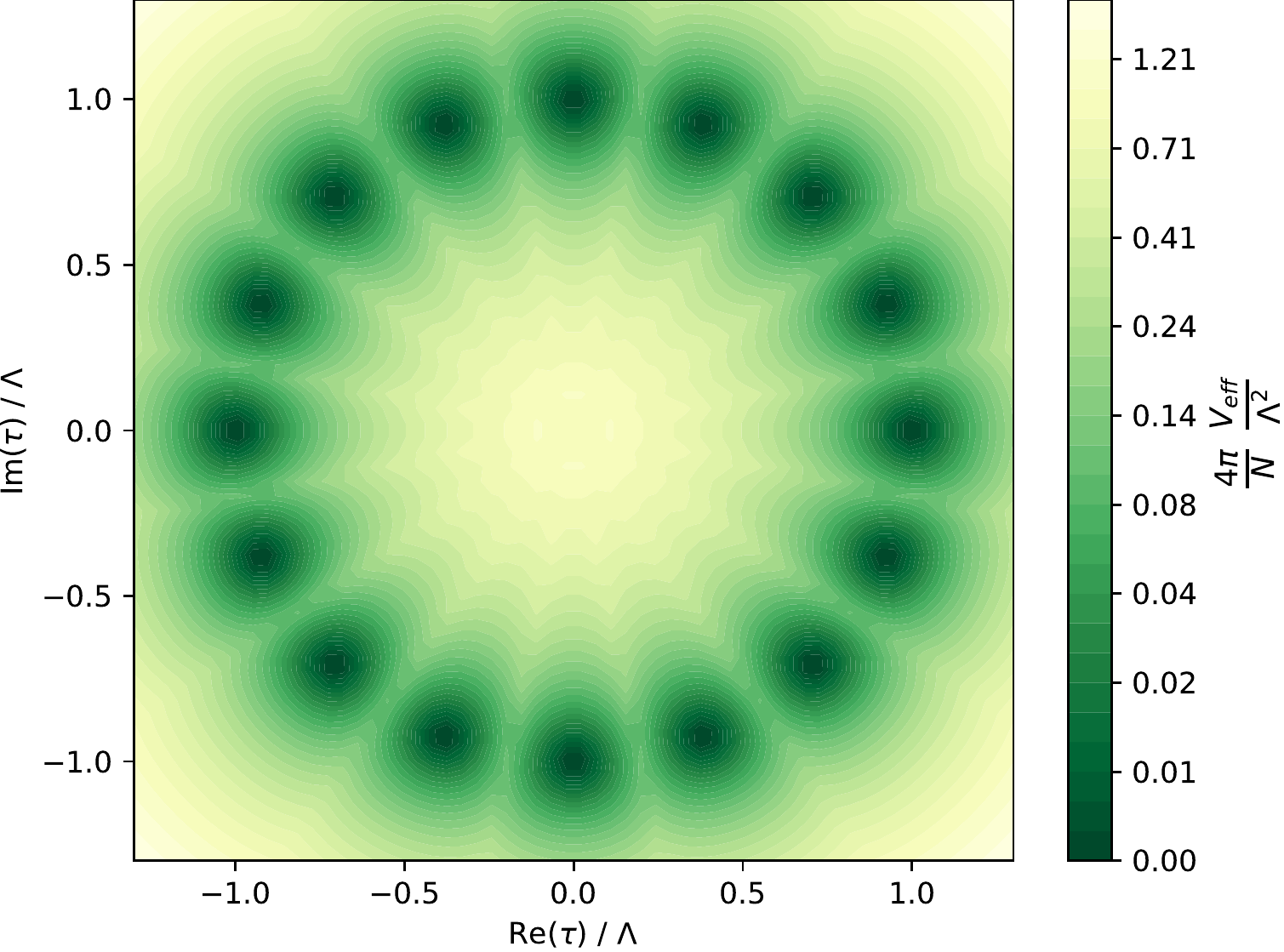}
        \caption{$\lambda = 0$, supersymmetric case, degenerate vacua}
        \label{fig:v_complex_tau_degenerate}
    \end{subfigure}%
    ~ 
    \begin{subfigure}[t]{0.5\textwidth}
        \centering
        \includegraphics[width=\textwidth]{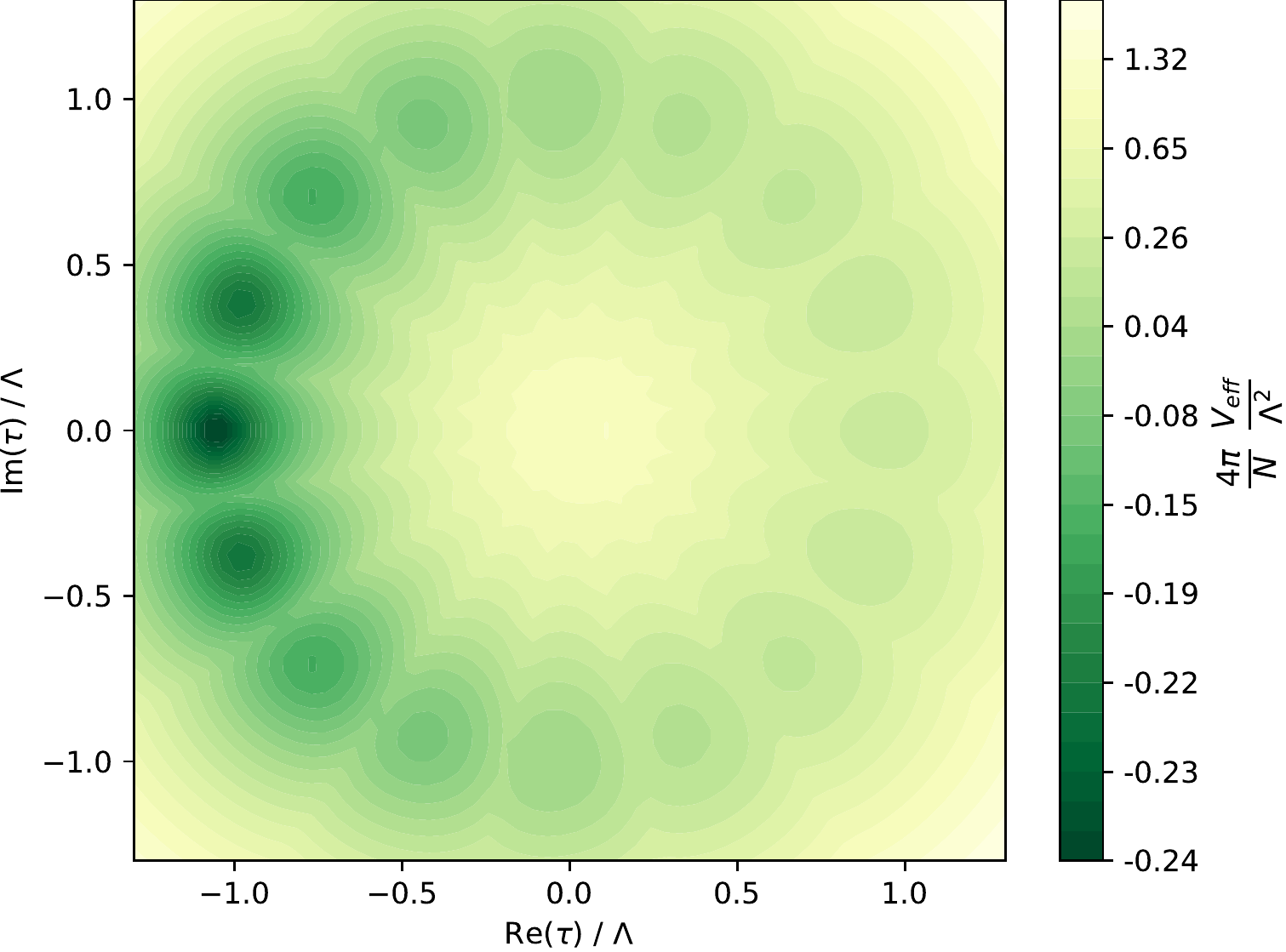}
        \caption{$\lambda > 0$, broken supersymmetry, lifted quasivacua}
        \label{fig:v_complex_tau_lifted}
    \end{subfigure}%
\caption{Effective potential \eqref{V_eff_strong_dm=0} on the complex $\tau = \sqrt{2}\sigma - m_0$ plane, with $D$ integrated out.}
\label{fig:v_complex_tau}
\end{figure}

\subsection{Equal mass case, small deformations}

We start by investigating the simplest case of equal mass parameters,
\begin{equation}
	m_0 = m_1 = \ldots = m_{N-1} \equiv m
\label{equal_masses}	
\end{equation}
Under this assumption  the potential proportional to  $\upsilon (\mu)$ is zero and the only deformation  we are left with is the fermion mass $\lambda$. For now we will not write its dependence on $\wt{\mu}$ explicitly. 


To simplify the equations, let us denote
\begin{equation}
	\tau = \sqrt{2}\sigma - m_0
\end{equation}
%
Then  the effective potential becomes
\begin{equation}
\begin{aligned}
	{\mathcal  V}_\text{eff} 
			&= \frac{N}{4\pi} iD\left[ 1 - \ln\frac{iD + \bigl| \tau \bigr|^2}{\Lambda^2} \right] 
			+ \frac{N}{4\pi} \bigl| \tau \bigr|^2 \left[ 1 -  \ln\frac{iD  + \bigl| \tau \bigr|^2}{m_G^2} \right] \\
			&-  \frac{N}{4\pi} \bigl| \tau - \lambda (\mu) \bigr|^2
				\left[ 1 - \ln\frac{\bigl| \tau - \lambda (\mu) \bigr|^2}{m_G^2} \right] + \Delta V(\arg\tau),
			\end{aligned}
\label{V_eff_strong_dm=0}	
\end{equation}
where $\tau = |\tau|\,e^{i \arg\tau}$. Here we added  a new term $\Delta V(\arg\tau)$ absent in \eqref{v_eff}. 
It takes into account the chiral anomaly and appears already in \ntwot \CP model at $\wt{\mu}=0$. As was shown by Witten
\cite{W79} the photon become massive due to the chiral anomaly with mass equal $2\Lambda$ . The complex scalar 
$\sigma$ is a superpartner of the photon and also acquires mass $2\Lambda$. In particular, its argument $\arg\tau$
becomes massive.

 This effect is taken into account  by the additional potential $\Delta V(\arg\tau)$ in \eqref{V_eff_strong_dm=0}. It is constructed as follows.  At small $\wt{\mu}$ VEVs of $\tau$ are approximately given by their
supersymmetric values,
\begin{equation}
	\tau^\text{SUSY}_k = - \Lambda \exp \left( \frac{2\pi\,i\, k}{N}	\right), \quad k=0, ..., N-1,
\label{vac_tau_susy_dm=0}	
\end{equation}
cf. \eqref{22sigmaapp}. 
We divide $2\pi$ into $N$ patches centered at vacuum values, $\arg\tau^\text{SUSY}_k=2\pi k/N +\pi$,
$k=0,...,(N-1)$ and define the potential $\Delta V(\arg\tau)$ to be quadratic in each patch. Namely, we have
\beq
\Delta V(\arg\tau) = \frac{N}{4\pi} \frac{m_{\arg\tau}^2}{2} (\arg\tau - \arg\tau^\text{SUSY}_k)^2,
\qquad \frac{2\pi (k-\frac12)}{N} < \arg\tau -\pi < \frac{2\pi (k+\frac12)}{N},
\label{DeltaV}
\eeq
where $m_{\arg\tau}$ is the mass of $\arg\tau$. We present its calculation in Appendix, in particular showing corrections (see eq. \eqref{m_arg_sigma_leading})
to the Witten's result \cite{W79}
\beq
m_{\arg\tau}^\text{SUSY} =2\Lambda.
\label{wittenmass}
\eeq

Without the additional potential $\Delta V(\arg\tau) $ $N$ discrete vacua \eqref{vac_tau_susy_dm=0} disappear immediately
as we switch on $\wt{\mu}$ due to the lifting of quasivacua. We show below that with $\Delta V(\arg\tau) $ taken into account quasivacua are 
still present at small $\wt{\mu}$ and disappear only at certain finite critical $\wt{\mu}_\text{crit}$ which we identify as a phase transition point. Note that possible higher corrections to the quadratic potential \eqref{DeltaV} are suppressed in 
the large $N$ limit because the width of each patch is small, proportional to $1/N$.

\subsubsection{Vacuum energies}

As we turn on the deformation parameter $\wt{\mu}$ the  mass of $\xi^i$ fermion $\lambda(\wt{\mu})$ is no longer zero.
This breaks explicitly both  chiral symmetry and two-dimensional supersymmetry. As a result the $Z_N$ symmetry is broken and VEVs of  $\sigma$ are no longer located at a circle. 
Moreover, 
at  $\wt{\mu} = 0$  our model has $N$ degenerate vacua given by \eqref{vac_tau_susy_dm=0}.
When we switch on $\wt{\mu}$,  the corresponding vacuum energies split, and all vacua except the one at $k=0$  become quasivacua. The only true vacuum is the one at $k=0$, see Fig. \ref{fig:v_complex_tau}. As we discussed in Sec.~1.1 this leads to the confinement of kinks.

It turns out that there are two mechanisms responsible for the vacuum energy splitting. One is due to the effective potential \eqref{V_eff_strong_dm=0} and dominates at small $\wt{\mu}$.  The other one is typical for the non-supersymmetric
\CP model, see Sec.~1.1. It is  due to the constant electric field of  kinks interpolating between neighboring quasivacua
and dominates at large $\wt{\mu}$. 
We will now study the former mechanism, while the latter one will be considered in the next subsection.

Energy splittings in the small $\wt{\mu}$ limit  can be derived using the small $\lambda (\mu)$ expansion of the effective potential \eqref{V_eff_strong_dm=0}:
\begin{equation}
	{\mathcal  V}_\text{eff} = {\mathcal  V}_\text{SUSY}  + \delta {\mathcal  V},
\end{equation}
where ${\mathcal  V}_\text{SUSY}$ is the supersymmetric effective potential corresponding to $\lambda = 0$, while
\begin{equation}
	\delta {\mathcal  V} \approx \frac{N}{4\pi} \cdot 2  \Re\tau \cdot \lambda  \, \ln\frac{m_G^2}{\bigl| \tau \bigr|^2}
\label{dV_small-lam}	
\end{equation}
is the $O(\lambda)$ deformation. We can immediately infer lifted vacuum energies by plugging unperturbed VEVs \eqref{vac_tau_susy_dm=0}	into \eqref{dV_small-lam}. As we already mentioned the  ground state (true vacuum) is located at 
\beq
\tau_0 = - \Lambda = \Lambda e^{i \pi}, 
\label{yruevacuum}
\eeq
while  the first quasivacuum is at 
\begin{equation}
	\tau_1 = - \Lambda \exp \left(\frac{2\pi\,i}{N} \right) \approx - \Lambda - \Lambda \frac{2\pi\,i}{N} + \Lambda \frac{2\pi^2}{N^2}
\label{first_quasivac_strong}	
\end{equation}
Plugging this into \eqref{dV_small-lam} we get for the vacuum splitting\footnotemark
\begin{equation}
	E_1 - E_0 = \frac{2 \pi}{N} \lambda\Lambda  \ln \frac{m_G}{\Lambda}
\label{splitting_1-0}	
\end{equation}
\footnotetext{Formula \eqref{splitting_1-0} has a correction coming from the energy-momentum trace anomaly, but this correction is of the next order in the small parameter $\frac{\lambda}{\Lambda}  \ln \frac{M_\text{uv}}{\Lambda}$.}
This signifies that kinks interpolating between these vacua are now confined, as opposed to the supersymmetric case.

\subsubsection{Corrections to the VEVs}

\begin{figure}[h]
    \centering
    \begin{subfigure}[t]{0.5\textwidth}
        \centering
        \includegraphics[width=\textwidth]{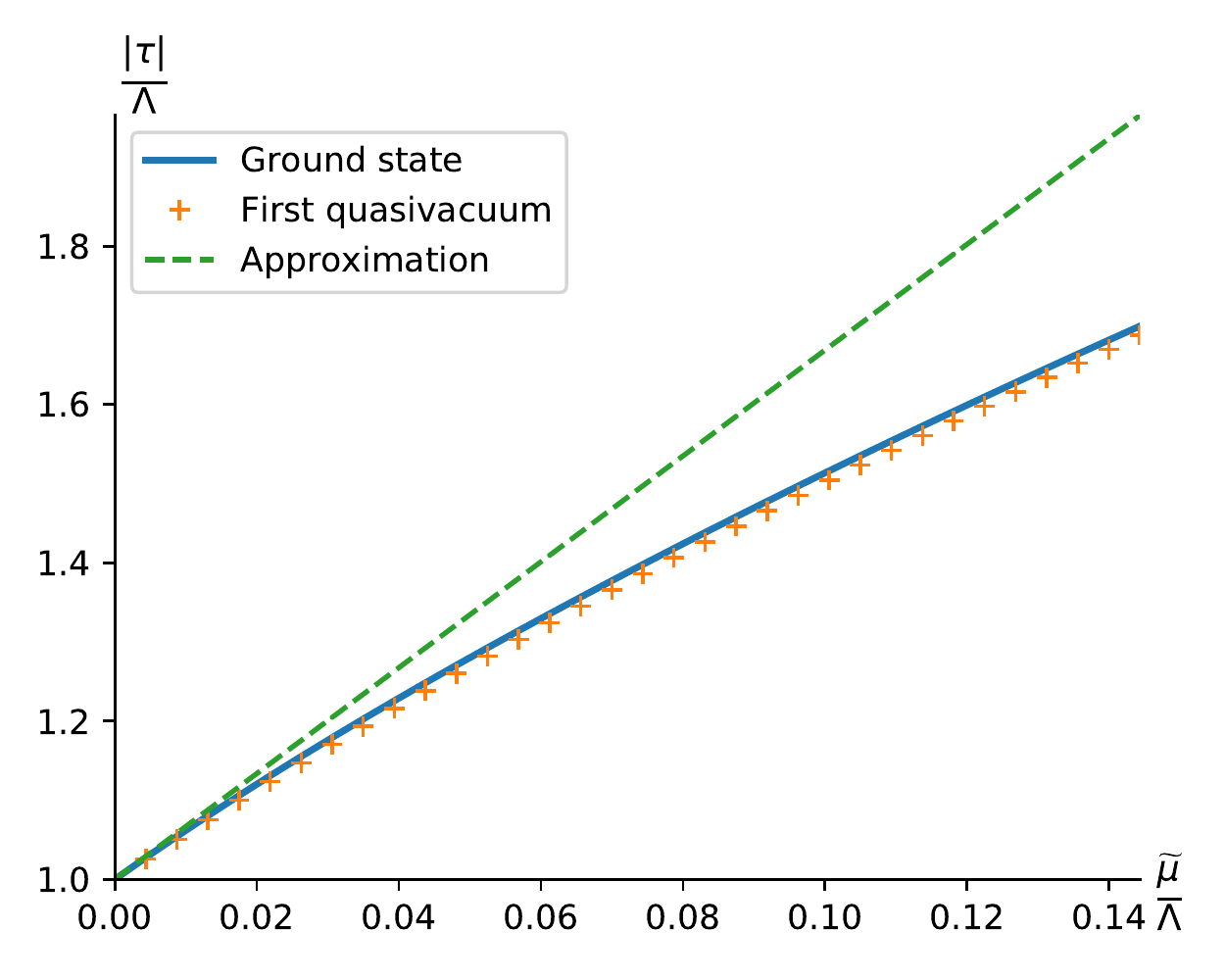}
        \caption{$|\tau_\text{ground}|$ and $|\tau_1|$}
        \label{fig:dm=0_smalllam_tau_abs}
    \end{subfigure}%
    ~ 
    \begin{subfigure}[t]{0.5\textwidth}
        \centering
        \includegraphics[width=\textwidth]{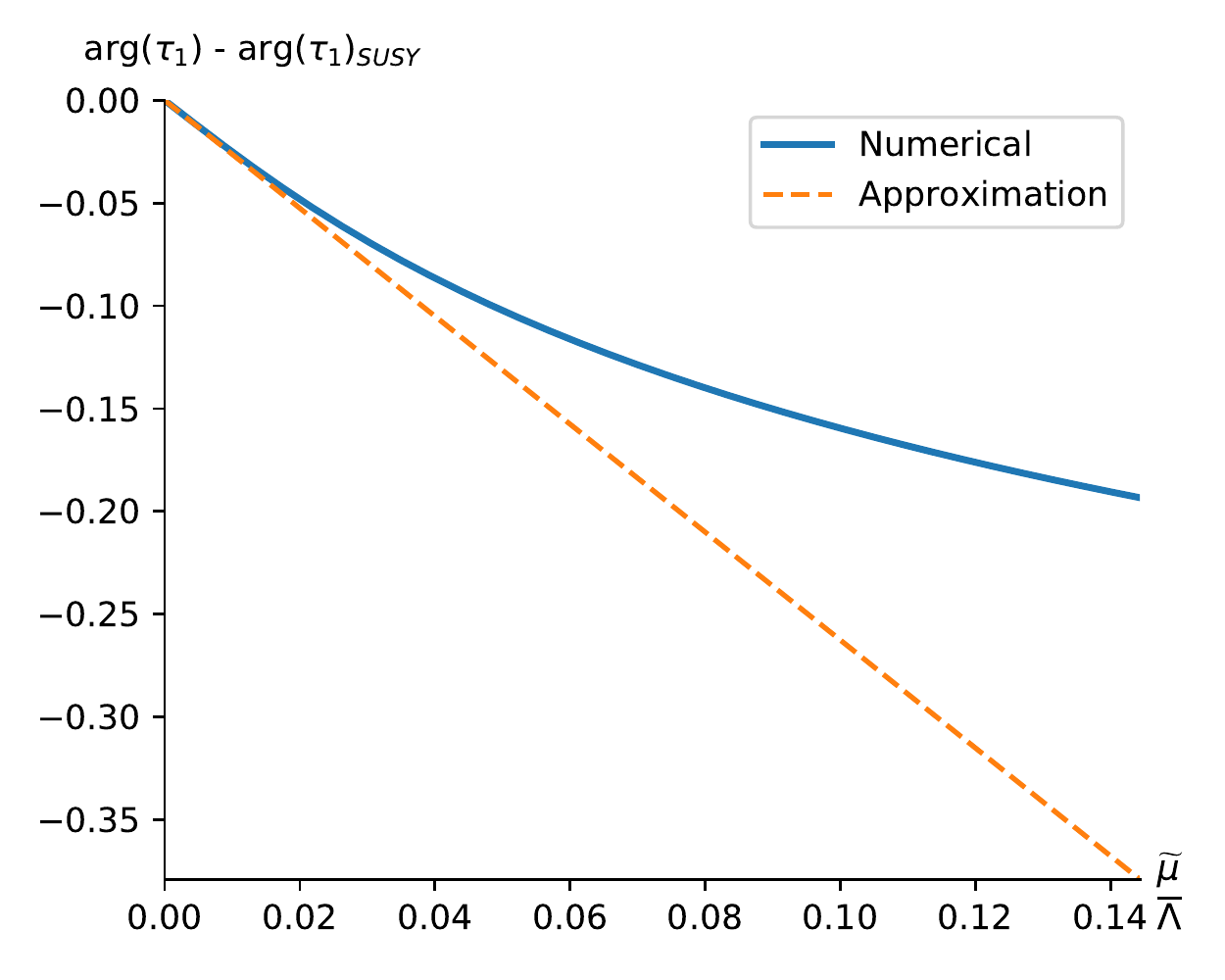}
        \caption{Correction to $\arg\tau_1$}
        \label{fig:dm=0_smalllam_tau_arg}
    \end{subfigure}%
\caption{
	Numerical results for the minima $\tau_\text{ground}$ and $\tau_1$ obtained by directly minimizing \eqref{V_eff_strong_dm=0}. On the figure \subref{fig:dm=0_smalllam_tau_abs}, the green dashed line shows the approximate formula \eqref{ground_state_correction}, the solid blue line is the numerical values of $|\tau_\text{ground}|$, while $|\tau_1|$ is shown by red \textquote{+}. 
	Figure \subref{fig:dm=0_smalllam_tau_arg} shows the approximate correction to $\arg\tau_1$ (\eqref{first_quasivacuum_correction}, the last term) and the numerical results for this quantity. 
}
\label{fig:dm=0_smalllam_tau}
\end{figure}

Now let us derive corrections to the unperturbed VEVs \eqref{vac_tau_susy_dm=0}.
Minimizing the potential \eqref{V_eff_strong_dm=0} we get:
\begin{equation}
	2\beta_\text{ren} = \ln\frac{ iD + \bigl| \tau \bigr|^2}{\Lambda^2} = 0 \Rightarrow  iD + \bigl| \tau \bigr|^2 = \Lambda^2
\label{master1_dm=0}	
\end{equation}
\begin{equation}
	|\tau| \ln\frac{\bigl| \tau - \lambda (\wt{\mu}) \bigr|^2}{\Lambda^2} + \cos(\arg\tau) \lambda (\wt{\mu}) \ln\frac{m_G^2}{\Lambda^2} = 0
\label{stronf_dm=0_taueq}
\end{equation}
\begin{equation}
	-  \sin(\arg\tau) \, \lambda |\tau| \, \ln\frac{m_G^2}{\Lambda^2} + \frac{m_{\arg\tau}^2}{2} \, (\arg\tau - \arg\tau^\text{SUSY}_k) = 0
\end{equation}
The approximate solution  in the limit of small $\wt{\mu}$ is given by
\begin{equation}
	|\tau| \approx \Lambda - \cos(\arg\tau^\text{SUSY}_k) \frac{1}{2} \lambda \ln\frac{m_G^2}{\Lambda^2}
\end{equation}
\begin{equation}
	\arg\tau \approx \arg\tau^\text{SUSY}_k  + \sin(\arg\tau^\text{SUSY}_k) \, \frac{2\lambda\Lambda}{m_{\arg\tau}^2}  \, \ln\frac{m_G^2}{\Lambda^2} 
\end{equation}
In particular, for the $\tau_0 = - \Lambda$ we get the corrected value
\begin{equation}
	\tau_\text{ground} \approx - \Lambda - \frac{1}{2} \lambda \ln\frac{m_G^2}{\Lambda^2}
\label{ground_state_correction}	
\end{equation}
while for the first quasivacuum \eqref{first_quasivac_strong}
\begin{equation}
\begin{aligned}
	|\tau_1| &\approx |\tau_\text{ground}| \approx  \Lambda + \frac{1}{2} \lambda \ln\frac{m_G^2}{\Lambda^2} \\
	\arg\tau_1 &\approx \underbrace{ \left(\pi + \frac{2\pi}{N} \right) }_\text{unperturbed} 
		- \frac{2\pi}{N} \frac{\lambda}{2 \Lambda}  \, \ln\frac{m_G^2}{\Lambda^2} 
\end{aligned}	
\label{first_quasivacuum_correction}
\end{equation}
where we used \eqref{wittenmass} for the non-perturbed  mass of $\sigma$. These results agree with numerical calculations, see Fig. \ref{fig:dm=0_smalllam_tau}.

Note that when 
\begin{equation}
	\frac{\lambda}{\Lambda}  \, \ln\frac{m_G}{\Lambda} = 1
\label{tau_0=tau_1}	
\end{equation}
we have in our approximation $\arg\tau_1 = \tau_\text{ground} = \pi$, and the quasivacuum at $\tau_1$ effectively disappears. This signifies that around the point \eqref{tau_0=tau_1}	a phase transition might take place. This will turn out to be true, see Sec. \ref{sec:strong1_strong2} below.

The quasivacuum with the highest energy is located at
\begin{equation}
	\tau_\text{high} \approx  \Lambda - \frac{1}{2} \lambda \ln\frac{m_G^2}{\Lambda^2}
\end{equation}
Further analysis of the equation \eqref{stronf_dm=0_taueq} shows that this solution disappears at
\begin{equation}
	\lambda = \frac{2\Lambda}{e \ln \frac{m_G^2}{\Lambda^2}}
\label{strong1_strong2_trans_potential}	
\end{equation}
which is consistent with \eqref{strong1_strong2_trans_arg-tau}. This suggests that around the critical value of the deformation 
\beq
\lambda_\text{crit} \sim \frac{\Lambda}{ \ln \frac{m_G^2}{\Lambda^2}}
\label{strong1_strong2_trans_arg-tau}
\eeq
all quasivacua have decayed (cf. \eqref{tau_0=tau_1}).

\subsection{Effective action \label{sec:effact}}

As we already mentioned there are two mechanisms of the energy splitting of quasivacua at non-zero $\wt{\mu}$. 
Both lead to the confinement of kinks. The first one
is due to $\wt{\mu}$-corrections present in the effective potential  \eqref{V_eff_strong_dm=0} These corrections  lift 
$\sigma$-quasivacua and lead to the  splitting  described by Eq.~\eqref{splitting_1-0}. The second mechanism is due to the 
constant electric field of kinks interpolating between quasivacua. The photon $A_\mu$ becomes dynamical on the quantum level
\cite{W79}. 
We will see below that, as we turn on the deformation parameter $\wt{\mu}$, the photon acquires a massless component. A linear Coulomb potential is  generated, but the vacuum energy splitting due to the electrical field is much smaller then the one in 
\eqref{splitting_1-0}.
At sufficiently large $\wt{\mu}$ all $N-1$  $\sigma$-quasivacua decay, and the splitting is saturated by the electric field only.
We identify this change of the regime and associated discontinuity in (the derivative of) $(E_1-E_0)$  as a phase transition.

\subsubsection{Derivation of the effective action \label{sec:effact_deriv}}

Consider now the effective action of our $\wt{\mu}$-deformed \CP model \eqref{lagrangian_init} obtained by integrating out 
$n^i$ and $\xi^i$ fields in the large-$N$ approximation. Relaxing the condition that $\sigma$ and $D$ are constant fields
assumed in Sec.~2.1  we consider the one loop effective action as a functional of  fields of the vector supermultiplet.

Considering the vicinity of the true vacuum where $\Im\langle\sigma\rangle = 0$  we write down the bosonic part of the action
in the form (Minkowski formulation\footnotemark)
\footnotetext{In this subsection we will use the Minkowski formulation with $g^{\mu\nu} = \text{diag} \{+,-\}$, and for the Levi-Civita symbol $\varepsilon_{01} = - \varepsilon^{01} = +1$.}
%
\begin{equation}
	S_{\rm eff}=
	 \int d^2 x \left\{
	- \frac1{4e_{\gamma}^2}F^2_{\mu\nu} + \frac1{e_{\Im\sigma}^2}	|\pt_{\mu}\Im\sigma|^2 + \frac1{e_{\Re\sigma}^2}	|\pt_{\mu}\Re\sigma|^2
	- V(\sigma) - \sqrt{2} \, b_{\gamma,\Im\sigma} \, \Im\sigma \,F^{*}
	  \right\},
	\label{effaction}
\end{equation}
where 
$F^{*}$ is the  dual gauge field strength,
\begin{equation}
	F^{*}= - \frac12\varepsilon_{\mu\nu}F^{\mu\nu}\,.
\end{equation}
This effective action was first presented  for \ntwot and \ntwoo supersymmetric \CP models in \cite{SYhetN}.
Here we generalize it for the $\wt{\mu}$-deformed \CP model \eqref{lagrangian_init}. The potential 
 $V(\sigma)$ here can be obtained  from \eqref{v_eff} by eliminating $D$ by virtue of its equation of motion.

Coefficients in front of $A_{\mu}$ and $\sigma$ kinetic terms are  finite after renormalization reflecting
 Witten's observation  that these fields become physical \cite{W79}.
The last term in \eqref{effaction} is $A_\mu - \sigma$ induced by the chiral anomaly. Because of this mixing, the would-be massless photon and the phase of $\sigma$ acquire a mass \eqref{wittenmass} already in unperturbed theory at $\wt{\mu}=0$.
This term is also present when we switch on the deformation.

\begin{figure}[h!]
    \centering
    \begin{subfigure}[t]{0.33\textwidth}
        \centering
        \includegraphics[width=\textwidth]{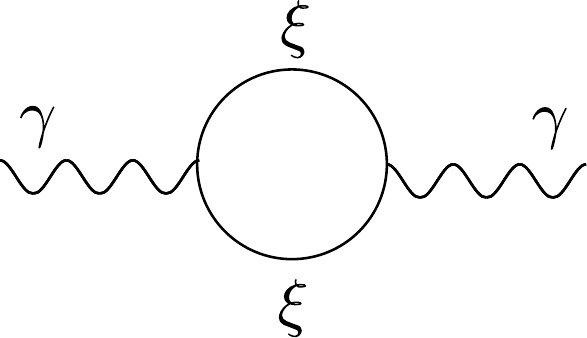}
        \caption{Photon wave function renormalization}
        \label{fig:loops:photon}
    \end{subfigure}%
    \hspace{20pt} %
    \begin{subfigure}[t]{0.33\textwidth}
        \centering
        \includegraphics[width=\textwidth]{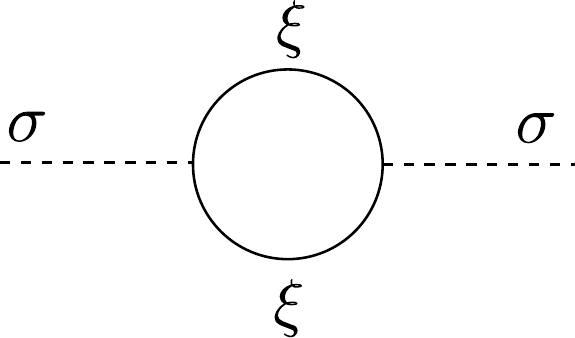}
        \caption{Scalar wave function renormalization}
        \label{fig:loops:scalar}
    \end{subfigure}
    
    \vspace{20pt}
    
    \begin{subfigure}[t]{0.33\textwidth}
        \centering
        \includegraphics[width=\textwidth]{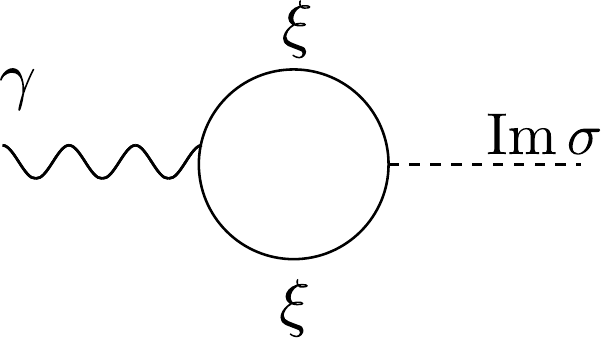}
        \caption{Photon-scalar mixing}
        \label{fig:loops:mixing}
    \end{subfigure}
\caption{Contributions to the effective action}
\label{fig:loops}
\end{figure}

Coefficients in this effective action come from loops. We take the low-energy limit when the external momenta are small. There are several contributions.
Photon wave function renormalization comes from the diagram on Fig. \ref{fig:loops:photon} and a similar graph with a bosonic loop. 
Wave function renormalizations for $\Re\sigma$ and $\Im\sigma$ come from the diagram on Fig. \ref{fig:loops:scalar} and also similar graph with a bosonic loop. Finally, the mixing term is given by the diagram on Fig. \ref{fig:loops:mixing}. For the mass distribution \eqref{masses_ZN} and the vacuum with $\Im\langle\sigma\rangle = 0$, the normalization factors are:
%
%
\begin{equation}
\begin{aligned}
	\frac{1}{e^2_{\Re\sigma}} &= \frac{1}{4\pi}\,\sum_{k=0}^{N-1} 
		\left[ \frac{1}{3} \frac{M_{\xi_k}^2 + 2 \left( \Im m_k \right)^2 }{M_{\xi_k}^4} 
			+ \frac{2}{3} \frac{\left( \sqrt{2}\langle\sigma\rangle - \Re m_k \right)^2}{m_{n_k}^4} \right]
	\\
	\frac{1}{e^2_{\Im\sigma}} &= \frac{1}{4\pi}\,\sum_{k=0}^{N-1} 
			\left[ \frac{1}{3} \frac{3 M_{\xi_k}^2 - 2 \left( \Im  m_k \right)^2 }{M_{\xi_k}^4} 
				+ \frac{2}{3} \frac{\left( \Im m_k \right)^2}{m_{n_k}^4} \right]
	\\
	\frac{1}{e^2_{\gamma}} &= \frac{1}{4\pi}\,\sum_{k=0}^{N-1} \left[ \frac{1}{3} \frac{1}{m^2_{n_k}} + \frac{2}{3} \frac{1}{M^2_{\xi_k}} \right]
	\\
	b_{\gamma,\Im\sigma} &= \frac{1}{2\pi}\,\sum_{k=0}^{N-1} \frac{ \sqrt{2}\langle\sigma\rangle - m_k - \lambda (\wt{\mu}) }{M_{\xi_k}^2}
\end{aligned}
\label{eff_normalizations}
\end{equation}
Here, $M^2_{\xi_k}$ and $m^2_{n_k}$ are the masses of the $\xi_k$ and $n_k$ fields respectively:
\begin{equation}
\begin{aligned}
	M^2_{\xi_k} &= | \sqrt{2}\langle\sigma\rangle - m_k - \lambda (\wt{\mu}) \bigr|^2
	\\
	m^2_{n_k} &= i\langle D \rangle + \upsilon (\wt{\mu})\Delta m_k  + \bigl| \sqrt{2}\langle\sigma\rangle - m_k \bigr|^2
\end{aligned}
\end{equation}
We present details of this calculation in Appendix A.

Next we  diagonalize the photon-$\sigma$ mass matrix  in \eqref{effaction}, see  Appendix B. As we already mentioned this diagonalization
shows that the photon acquires a massless component as soon as we switch on $\wt{\mu}$. This component is responsible for the 
presence of the constant electric field in quasivacua. This constant electric field gives rise to a second mechanism of quasivacua splitting, see \eqref{Evac_electical-1}. This effect is small at small $\wt{\mu}$ but becomes dominant  at larger $\wt{\mu}$ above the phase transition
point. This result can also be derived in a different way which we consider in the next subsection.

\subsubsection{Coulomb potential and vacuum energies}

In this section we study the formation  of a constant electric field in a quasivacuum generalizing a method 
developed by Witten in \cite{W79} for \ntwot supersymmetric \CP model.

 Let us start with the effective action \eqref{effaction} taking into account the presence of the trial matter charges,
\begin{equation}
	S_{\rm eff}=
	 \int d^2 x \left\{
	- \frac1{4e_{\gamma}^2}F^2_{\mu\nu} - \sqrt{2} \, b_{\gamma,\Im\sigma} \, \Im\sigma \,F^{*}
	+ j_\mu A^\mu
	  \right\},
\end{equation}
Consider a stationary point-like kink at $x=x_0$ with electric charge $+1$ described by the current $j_{\mu} = (\delta(x-x_0),\, 0)$
and $F^{*} = -\frac12\varepsilon^{\mu\nu}F_{\mu\nu} = \p_0 A_1 - \p_1 A_0$. 

 We have the equation of motion for the photon:
\begin{equation}
	 - \frac{1}{e_\gamma^2} \p_x \mathcal{E} - \sqrt{2} \, b_{\gamma,\Im\sigma} \, \p_x \Im\sigma = - j_0,
\end{equation}
where
\begin{equation}
	\mathcal{E} = F_{01}  
\end{equation}
 is the electric field strength.
Integrating over the spatial coordinate we obtain
\begin{equation}
	  \frac{1}{e_\gamma^2} ( \mathcal{E}(\infty) - \mathcal{E}(-\infty)) + \sqrt{2} \, b_{\gamma,\Im\sigma} \, ( \Im\sigma(\infty) - \Im\sigma(-\infty)  )
	 = 1
\label{kink_topological_eq}	 
\end{equation}
In the supersymmetric case $\wt{\mu} = 0$ the photon is massive, so there is no constant electric field,  $\mathcal{E}(\infty) = \mathcal{E}(-\infty) = 0$. Therefore we have
\begin{equation}
	   \sqrt{2} \, b_{\gamma,\Im\sigma} \, ( \Im\sigma(\infty) - \Im\sigma(-\infty)  ) = 1
\end{equation}
Since 
\begin{equation}
	b_{\gamma,\Im\sigma} = \frac{1}{2\pi} \frac{N}{\Lambda},
\label{b_susy}	
\end{equation}
see Eq.~\eqref{eff_normalizations} for $\wt{\mu}=0$
we get
\begin{equation}
	\sqrt{2}  ( \Im\sigma(\infty) - \Im\sigma(-\infty)  ) = 2 \pi \frac{\Lambda}{N}
\end{equation}
which, if we set $ \tau (-\infty) = - \Lambda$ for the true vacuum, is an approximation of
\begin{equation}
	\tau(\infty) = - \Lambda e^{\frac{2\pi i}{N}}
\end{equation}
for the value of $\sigma$ VEV in the first quasivacuum, see \eqref{first_quasivac_strong}.
This result for the \ntwot case has been derived long ago by Witten \cite{W79} showing the presence of $N$ vacua and 
kinks interpolating between them. 

Now, consider small deformations in the Eq.~\eqref{kink_topological_eq} for  a kink interpolating between the ground state \eqref{ground_state_correction} at $x=- \infty$ and the first quasivacuum
\eqref{first_quasivacuum_correction} at $x=+ \infty$. Setting $\mathcal{E}(-\infty) = 0$  we get from 
\eqref{kink_topological_eq}	
\begin{equation}
	  \frac{1}{e_\gamma^2}  \mathcal{E}(\infty) + \sqrt{2} \, b_{\gamma,\Im\sigma} \, ( \Im\sigma(\infty) - \pi  ) = 1 
\end{equation}
Using \eqref{first_quasivacuum_correction} and \eqref{b_susy} we obtain for the electric field strength
\begin{equation}
	  \mathcal{E}(\infty) = e_\gamma^2 \, \frac{\lambda}{\Lambda}   \ln\frac{m_G}{\Lambda}. 
\end{equation}
We see that the kink  produces a constant electric field now.
This gives the contribution to the   energy density splitting between  the first quasivacuum and the true vacuum
\begin{equation}
	(E_{1} -E_0)|_{\mathcal{E}}  = \frac{1}{2 e_\gamma^2} \mathcal{E}^2 = \frac{2\pi}{N} \left(\lambda \ln \frac{m_G}{\Lambda} \right)^2
\label{Evac_electical-2}	
\end{equation}
This coincides with the result \eqref{Evac_electical-1} obtained from the photon-$\sigma$ diagonalization. This contribution is small compared to the $\sigma$-splitting   given by \eqref{splitting_1-0} at small $\wt{\mu}$.

\subsection{Second order phase transition \label{sec:strong1_strong2}}

\begin{figure}[h!]
	\centering
	\includegraphics[width=0.95\linewidth]{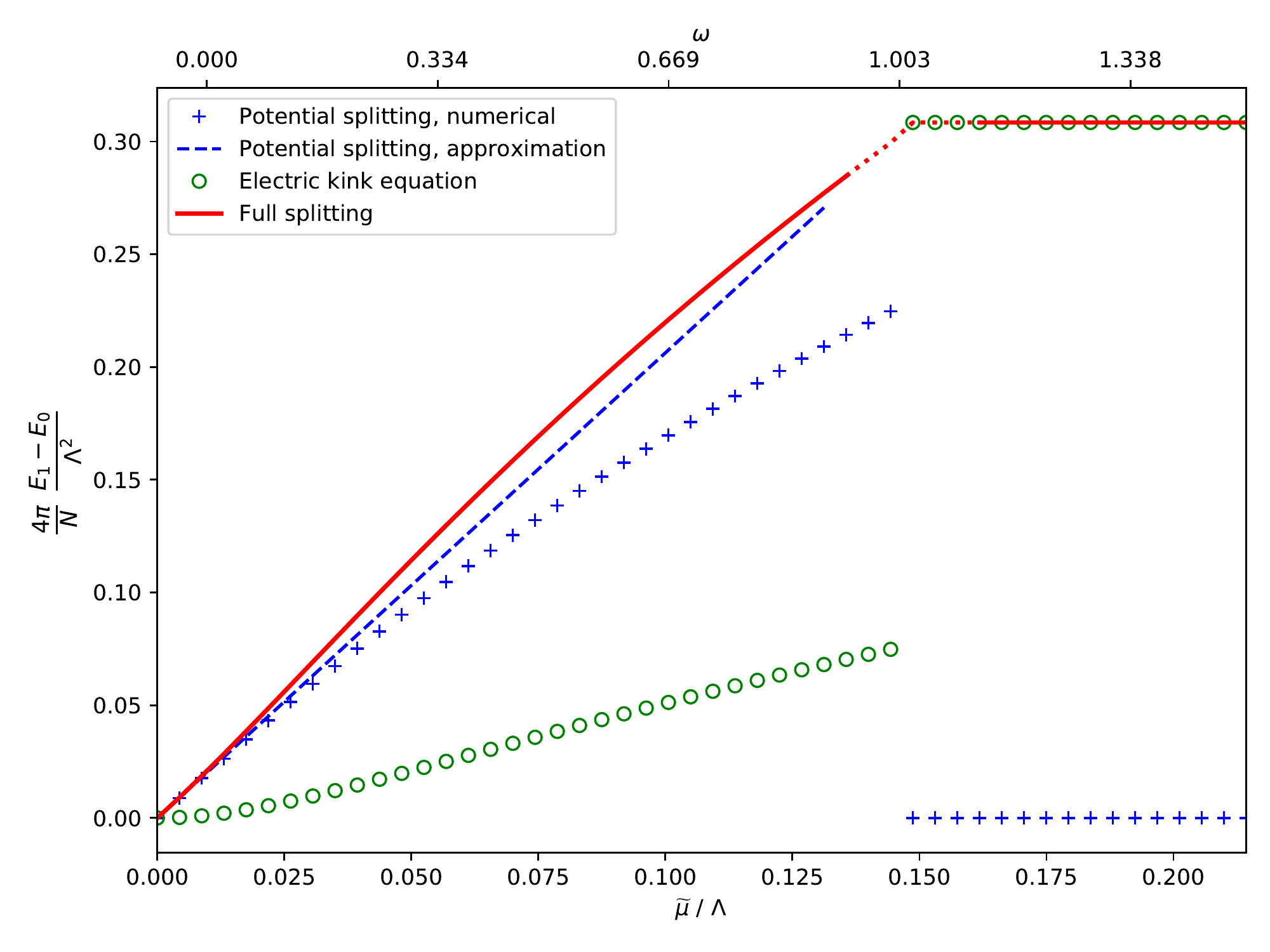}
	\caption{Different contributions to the vacuum energy. Vertical axis is labeled by the rescaled energy splitting $E_1 - E_0$. Values of the deformation parameter $\wt{\mu}$ are on the lower horizontal axis (in the units of $\Lambda$), while the upper horizontal axis represents the parameter $\omega$ \eqref{omega_splitting_parameter}. 
	Green circles denote the contribution from the electric field (solution of \eqref{kink_topological_eq}, given by \eqref{Evac_electical-2} below the phase transition point), \textquote{+} signs represent the splitting from the potential \eqref{V_eff_strong_dm=0} (the blue dashed line is the approximation \eqref{splitting_1-0}). The solid red line is the sum of these two contributions.	
	Phase transition occurs at $\omega \approx 1$ where the full energy displays a discontinuity of the first derivative. Our model does not allow us to obtain exact results in the vicinity of the phase transition point, and we have to extrapolate from the left and from the right (red dotted line continuing the solid red curve).}
\label{fig:vac_splittings} 
\end{figure}

As we learned so far, the vacuum energy (or, rather, energy splitting between the ground state and the first quasivacuum) has two contributions, which depend on the parameter
\begin{equation}
	\omega = \frac{\lambda (\wt{\mu})}{\Lambda} \, \ln\frac{m_G}{\Lambda}
\label{omega_splitting_parameter}	
\end{equation}

The first contribution is the splitting of different quasiminima  $\sigma_i$ of the effective potential \eqref{splitting_1-0}. When we turn on $\omega$ (i.e. supersymmetry breaking parameter $\wt{\mu}$), this contribution at first grows linearly with $\omega$, and then drops to zero when the $\sigma$-quasiminima  disappear.

The second contribution comes from the electric field of charged kinks interpolating between the quasivacua, see \eqref{Evac_electical-1} and \eqref{Evac_electical-2}. This contribution at first grows as $\omega^2$, and at the point when the first $\sigma$ quasivacuum disappears, electric field jumps up\footnotemark  to saturate \eqref{kink_topological_eq}. 
\footnotetext{This jumping is not seen from the propagator considerations \eqref{Evac_electical-1} since it holds only perturbatively near the true vacuum and does not take into account the presence of $\sigma$-quasivacua.
}

The jumping point is the same for these two contributions and it is where a phase transition occurs. Corresponding critical value is $\omega_c \sim 1$, i.e. (cf. \eqref{strong1_strong2_trans_arg-tau} and\eqref{tau_0=tau_1})
\begin{equation}
	\lambda_\text{crit} = \lambda (\wt{\mu}_\text{crit}) \sim \frac{\Lambda}{\ln \frac{m_G^2}{\Lambda^2}} \,.
\label{strong1_strong2_trans_electric}	
\end{equation}

Full vacuum energy is the sum of these two contributions, and on general grounds we expect that it does not jump. Rather, its first derivative is discontinuous, and the phase transition must be of the second order. Numerical calculations confirm this, see Fig. \ref{fig:vac_splittings}. At the point where the quasivacuum disappears, the two contributions to the vacuum energy jump, and the magnitudes are just right for the total sum to stay continuous.
However, we must point out that we do not have enough accuracy for the detailed study of the vicinity of the transition point. The point is that we can trust our formula for the $\arg\tau$ potential \eqref{DeltaV}  only in the vicinities of the minima $\eqref{vac_tau_susy_dm=0}$, and we do not know the exact form of this potential in regions between any of two adjacent minima.

At small deformations, the main contribution to the vacuum energy is $\sigma$-quasivacua splitting \eqref{splitting_1-0}. After the transition point, vacuum energy is determined solely by the kink electric field. As we reviewed in Sec.~1.1 it is the electric field that is responsible for the quasivacuum energy splittings in the non-supersymmetric \CP model. This is consistent with our results, since at large $\wt{\mu}$ above  the phase transition point our model flows to the non-supersymmetric \CP model.

To conclude this section we note that  parameter $\omega$ relevant for the quasivacua splitting  is enhanced by the large logarithm $\ln{m_G/\Lambda \gg 1}$. Hence the phase transition point occurs   at  $\wt{\mu}_c \sim \lambda _c$ given by
\eqref{strong1_strong2_trans_electric}, much smaller than $\wt{\mu} \sim \Lambda$. These are even smaller values of 
$\wt{\mu}$ as compared to $m_G$ since we assume $m_G\gg \Lambda$ in order to keep the bulk theory at weak coupling.
At these small values of $\wt{\mu}$ we are way below the scale of adjoint matter decoupling in the bulk theory 
which occurs at $\wt{\mu} \gg m_G$. In particular, the scale $\Lambda $ of the world sheet theory is close to $\Lambda_{4d}$ rather than
to its large-$\wt{\mu}$ asymptotic values \eqref{Lam_2d}.

\subsection{Large deformations \label{sec:strong_dm=0_largelam}}

\begin{figure}[ht]
    \centering
    \begin{subfigure}[t]{0.5\textwidth}
        \centering
        \includegraphics[width=\textwidth]{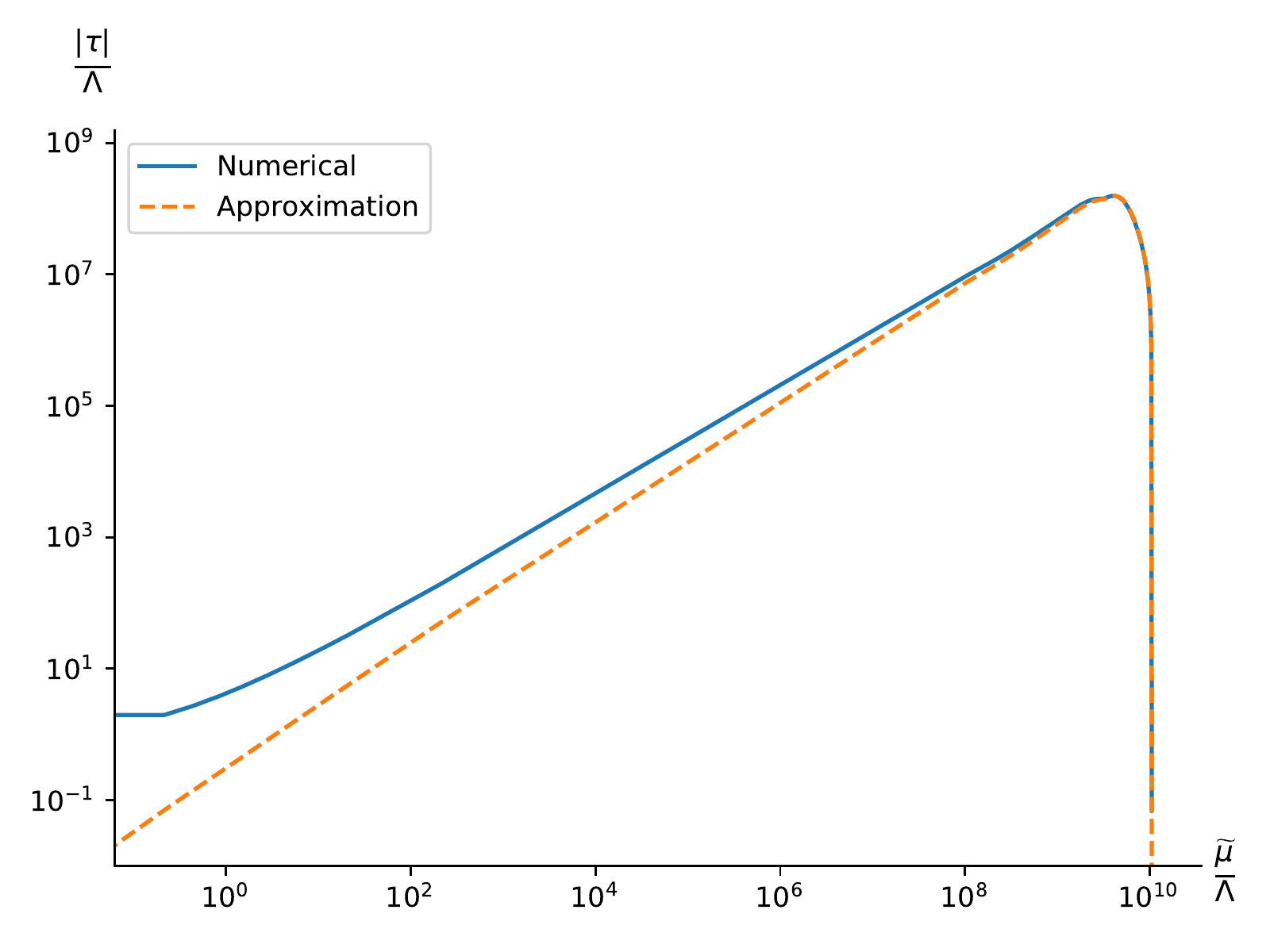}
        \caption{VEV of $\tau$ as a function of $\wt{\mu}$, double log scale. 
        	}
        \label{dm=0_large-lam_tau}  
    \end{subfigure}%
    ~ 
    \begin{subfigure}[t]{0.5\textwidth}
        \centering
        \includegraphics[width=\textwidth]{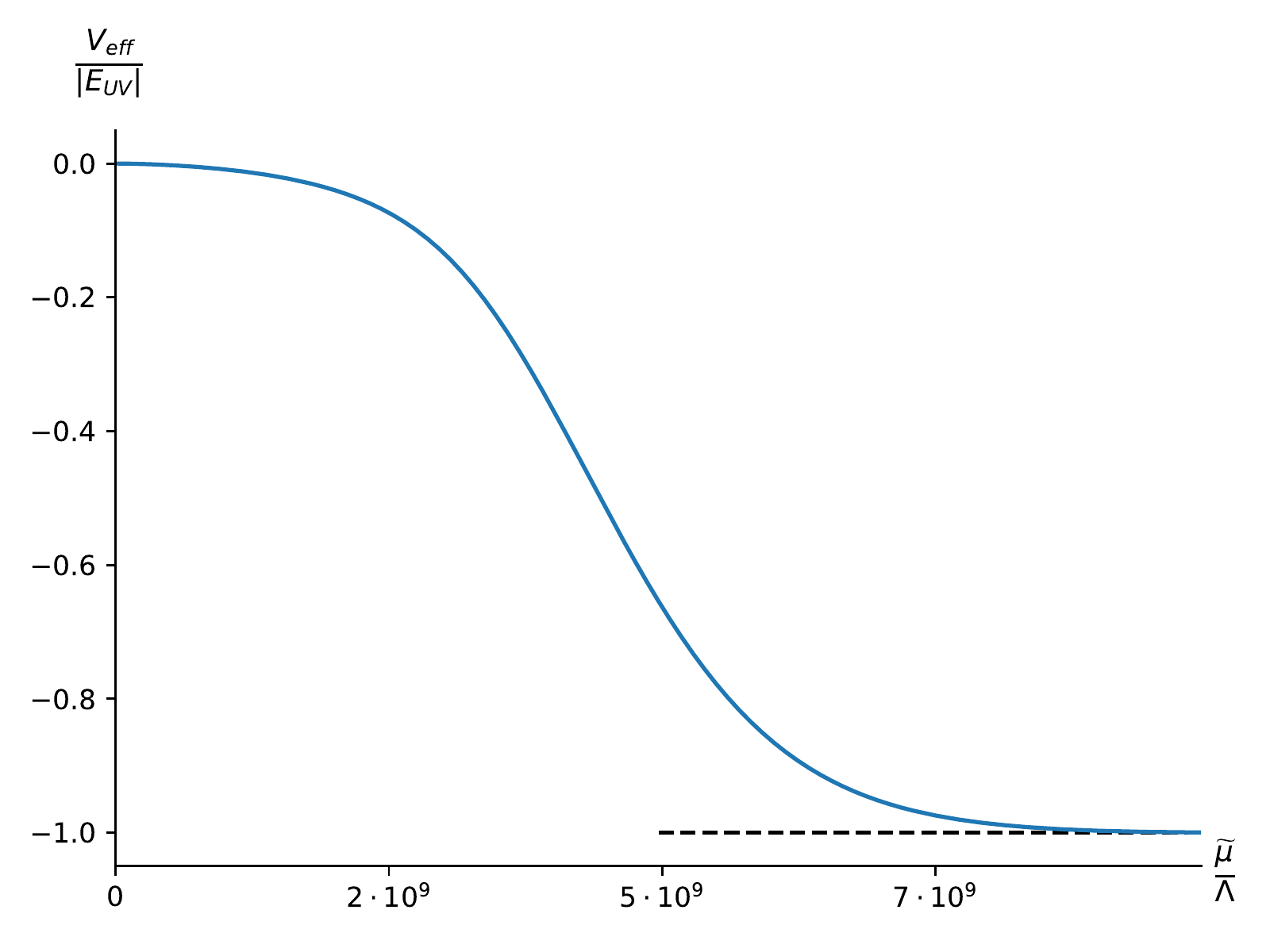}
        \caption{Vacuum energy as a function of $\wt{\mu}$, log scale}
	\label{dm=0_large-lam_Veff}          
    \end{subfigure}
    \caption{Numerical results for the VEV of $\tau$ and vacuum energy at large deformations $\lambda \gg 1$. 
    	On the figure \subref{dm=0_large-lam_tau} we have VEV of $\tau$. Dashed line shows the approximate solution \eqref{tau_approx}, while the solid line is the result of numerics. One can see that $\tau$ indeed vanishes at $\lambda(\wt{\mu}) = m_G$. 
    	On \subref{dm=0_large-lam_Veff} we have $E_\text{vac}$.  Dashed line shows its asymptotic value $E_\text{UV}$ given by \eqref{Evac_strong_largemu}.
    	In the numerical procedure we had set $m_G / \Lambda = 10^{10}$}
\label{dm=0_large-lam}    
\end{figure}

As we increase the  deformation parameter $\wt{\mu}$ , the fermion mass $\lambda$  approaches the UV cutoff scale $ m_G$ and  we can expect that the fermions become very heavy and decouple, effectively taking no part in the dynamics. Therefore, our theory should become the non-supersymmetric \CP model \eqref{cpn_lagr_simplest}. VEV of $\tau$ field should become zero.

  We can check this directly using our effective potential \eqref{V_eff_strong_dm=0}. Indeed, assume that $\tau \ll \lambda \sim m_G$. Then we can expand  \eqref{V_eff_strong_dm=0} to obtain
\begin{equation}
\begin{aligned}
	{\mathcal  V}_\text{eff} 
			&= \frac{N}{4\pi} iD\left[ 1 - \ln\frac{iD + \bigl| \tau \bigr|^2}{\Lambda^2} \right] 
			+ \frac{N}{4\pi} \bigl| \tau \bigr|^2 \left[ 1 -  \ln\frac{iD  + \bigl| \tau \bigr|^2}{m_G^2} \right] \\
			&-   \frac{N}{4\pi} \cdot 2  \Re\tau \cdot \lambda  \, \ln\frac{\lambda^2}{m_G^2}  	
			-  \frac{N}{4\pi} \lambda^2 \left( 1 - \ln\frac{\lambda^2}{m_G^2} \right) 
\end{aligned}
\label{V_eff_strong_dm=0_large-lam}	
\end{equation}
%
%
%
%
Minimizing this potential we obtain
\begin{equation}
	\tau \approx -\lambda \frac{\ln(m_G/\lambda)}{\ln (m_G/\Lambda) } \,.
\label{tau_approx}	
\end{equation}
This formula turns out to be pretty good compared to the exact numerical solution, see Fig. \ref{dm=0_large-lam_tau}.
As $\lambda$ approaches the UV cutoff scale $m_G$, the VEV of $\tau$ vanishes. The  first term in \eqref{V_eff_strong_dm=0_large-lam} reduces to the effective potential for the non-supersymmetric \CP model, while the last term gives a vacuum energy shift.  At $\lambda = m_G$, the vacuum energy is
\begin{equation}
	E_\text{vac, UV} = \frac{N}{4\pi} \left(\Lambda^2 - m_G^2 \right) \,.
\label{Evac_strong_largemu}
\end{equation}
This is in agreement with the Appelquist-Carazzone decoupling theorem \cite{Appelquist:1974tg}, which states that the effect of heavy fields is limited to the renormalization of physical quantities. Note that since the supersymmetry is explicitly 
broken in the world sheet theory by fermion masses the vacuum energy is not positively defined.

The vacuum energy  above is a quantum correction to the classical expression for the non-Abelian string tension in the bulk theory. The latter was derived in \cite{YIevlevN=1}, and together with \eqref{Evac_strong_largemu} it can be written as 
\begin{equation}
	T = \frac{2\pi}{\ln\frac{m_G^2}{m^2}} \frac{m_G^2}{g^2} + \frac{N}{4\pi} \left(\Lambda^2 - m_G^2\right),
\label{tension_strong}	
\end{equation}
 We see that the second term here is just an $O(g^2)$ correction to  the classical formula.

At intermediate values of $\lambda$ we were able to study this model only numerically. The results are presented on Fig. \eqref{dm=0_large-lam}. They show the dependence of $\langle \sigma \rangle$ and $E_\text{vac}$ on the heavy fermion mass $\lambda$. 
One can see that indeed the VEV of $\tau$ vanishes at very large $\lambda$.
Note that we will have $\langle iD \rangle < 0$ in a wide range of $\lambda$, but this does not lead to an instability because, according to \eqref{master1_dm=0}, the mass of the $n$ field is always positive.

%
%

\subsection{Split mass case \label{sec:strong_split_mass}}

\begin{figure}[h!]
	\centering
	\includegraphics[width=0.6\linewidth]{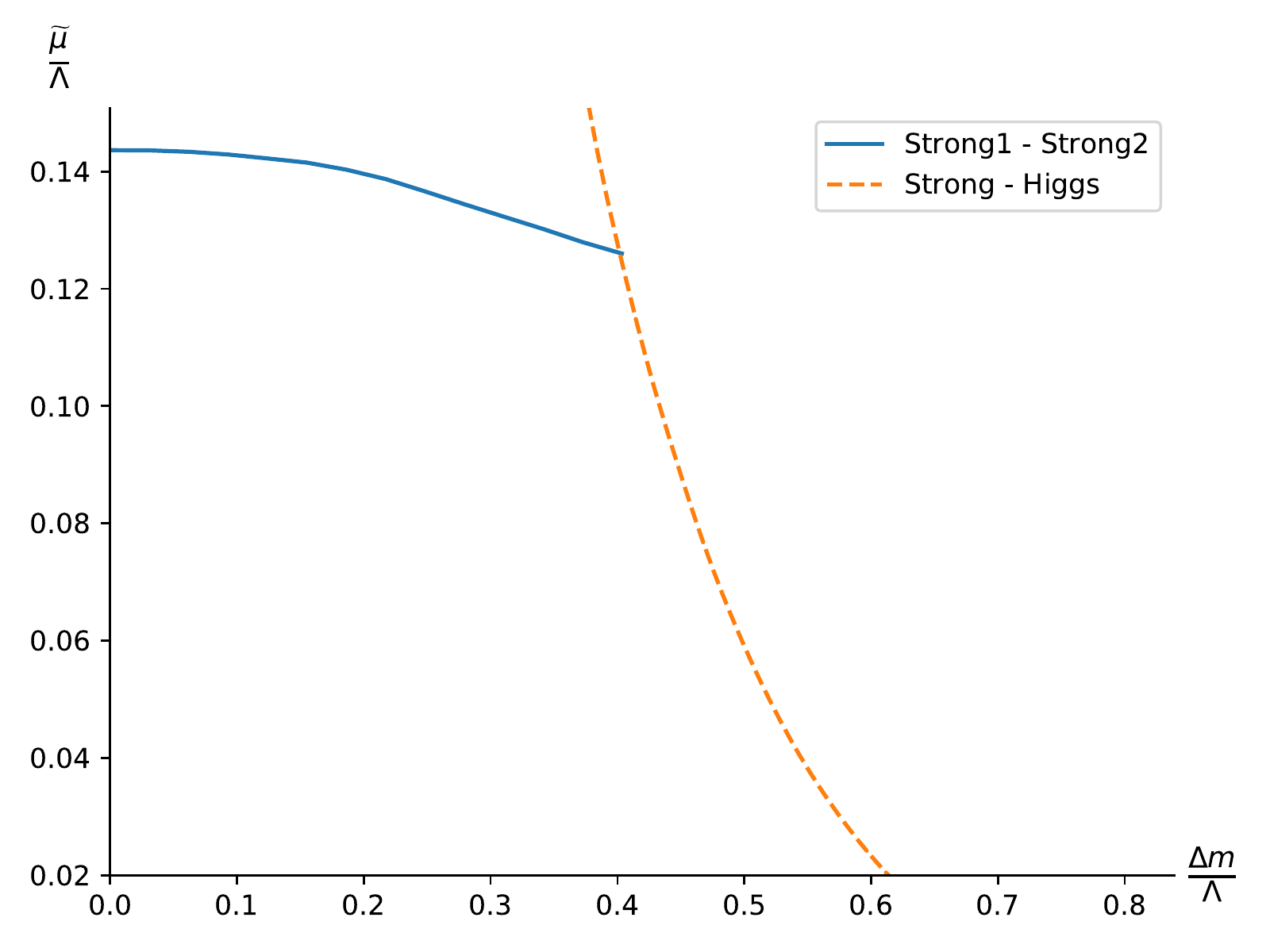}
	\caption{
		Phase transition line between two strong coupling regimes (shown in solid blue). The dashed line is the phase transition line between the Strong coupling and Higgs regimes, see Sec. \ref{sec:trans_strong-higgs}. This plot is a result of numerical calculations for $N = 16$.
	}
\label{fig:strong1_strong2} 
\end{figure}
The results obtained in the previous section can be generalized to the case $\Delta m_{i0} \neq 0$. Consider the masses on a circle \eqref{masses_ZN}, with the radius $\Delta m$ as the mass scale of our model.

If we fix some $\Delta m$ and start increasing $\wt{\mu}$ (and, therefore, $\lambda (\wt{\mu})$), our model exhibits similar behavior as in the case $\Delta m = 0$. At $\wt{\mu} = 0$ the supersymmetry is unbroken, and there are $N$ degenerate vacua. When we switch on the deformation, the degeneracy is lifted, and eventually all lifted quasivacua decay, which signifies a phase transition.
The set of the phase transition points represents a curve on the $(\mu, \Delta m)$ plane, see Fig. \ref{fig:strong1_strong2} 

Qualitatively, we see nothing new. However, when $\Delta m$ is large enough, the theory goes through the phase transition from the strong coupling phase into a weak coupling phase, so-called \textquote{Higgs} phase. This will be the subject of the next section.

%
%

\section{Higgs regime \label{sec:higgs}}


When the mass difference $\Delta m$ exceeds   some critical value, the theory appears in the Higgs phase. This phase is characterized by a nonzero VEV of $n$. At very weak coupling, we can use the classical Lagrangian \eqref{lagrangian_init} to find the vacuum solution,
\begin{equation}
	n_0^2 = 2 \beta \,, \quad
	\sqrt{2} \sigma = m_0 \,, \quad
	iD = 0 \,.
\label{higgs_classical}	
\end{equation}
The vacuum energy is classically zero.

In the supersymmetric case $\wt{\mu} = 0$ the solution for $\sigma$  
is exact at large $N$. Moreover, at very large $\Delta m$ the coupling constant $1/\beta$ is small (frozen at the scale $\Delta m$)
and quantum corrections to the classical vacuum solution \eqref{higgs_classical} are small.

However, at nonzero $\wt{\mu}$ and for $\Delta m \gtrsim \Lambda$, things become more complicated, as we can no longer rely on the classical equations. Generally speaking, solution \eqref{higgs_classical} receives $\Lambda /\Delta m $ and $\wt{\mu} / \Lambda$ corrections. We have to work with the quantum equations \eqref{master1} - \eqref{master3}, and most of the results presented in this section were obtained from numerical calculations.

\begin{figure}[h]
	\centering
	\includegraphics[width=0.6\linewidth]{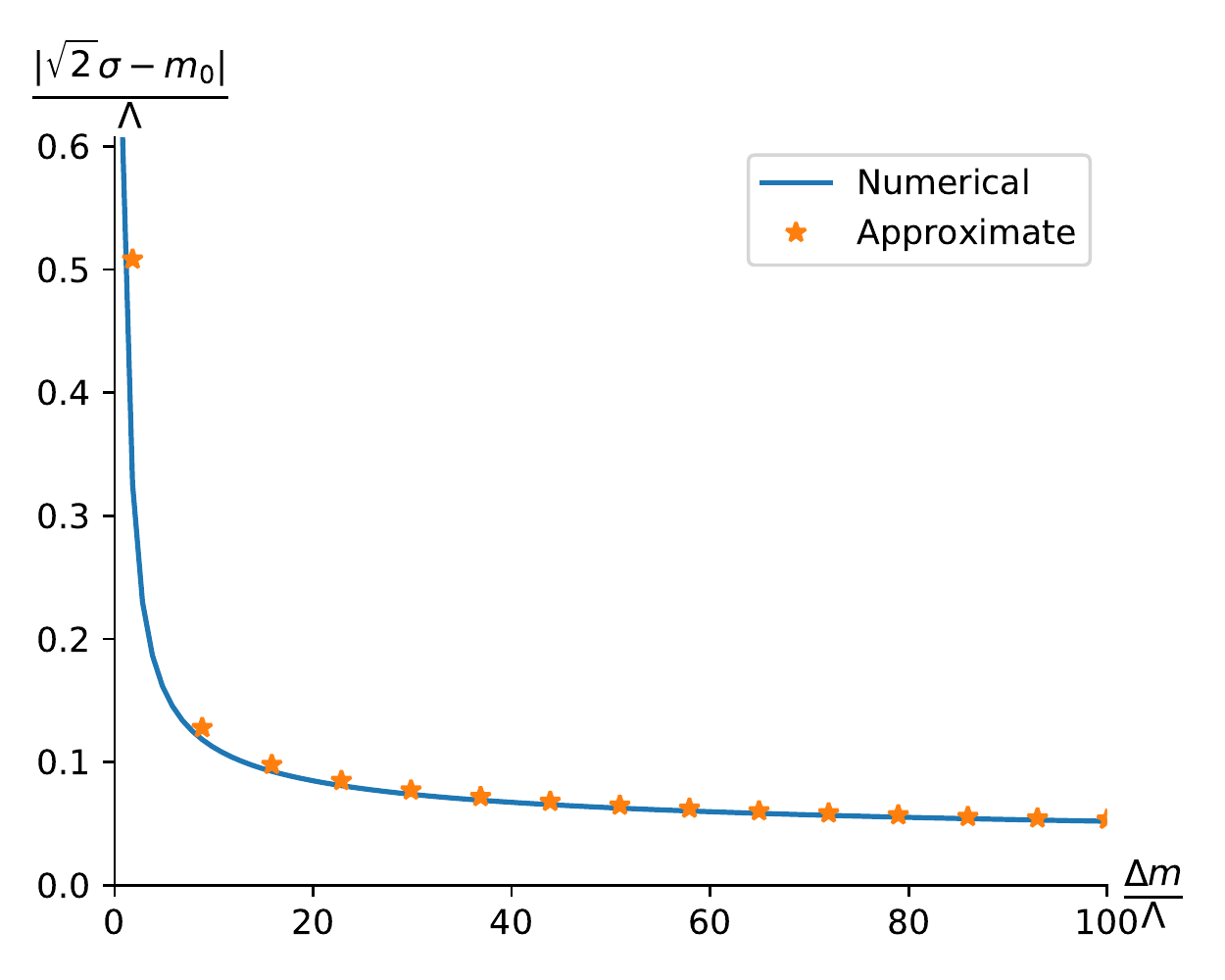}
	\caption{
		VEV of $\tau = \sqrt{2}\sigma - m_0$ as a function of $\Delta m$. 
		Solid line is the exact result of numerical calculation, while stars represent the approximate formula \eqref{higgs_tau_largedM}.
		Here $\wt{\mu} = \Lambda$. In numerical calculations we used $N = 16$.
		One can see that indeed, as $\Delta m$ grows, the VEV of $\sqrt{2}\sigma$ goes to its classical value $m_0$.
	}
\label{fig:quasiclassical_tauvev} 
\end{figure}
First of all, we wish to check that the one loop potential that we derived \eqref{v_eff} is compatible with the classical limit. Consider the limit of large $\Delta m \gg \Lambda$ with some $\wt{\mu}$ fixed. We can expand the vacuum equations \eqref{master1} - \eqref{master3} in powers of $\Lambda /\Delta m$ and easily derive an approximate solution for the ground state VEV
\begin{equation}
	\sqrt{2}\sigma - m_0 \approx - \lambda(\wt{\mu}) \, \frac{\ln \frac{m_G}{\Delta m}}{\ln\frac{\Delta m}{\Lambda}} \,.
\label{higgs_tau_largedM}	
\end{equation}

Fig. \ref{fig:quasiclassical_tauvev} presents our results for the VEV of $\sigma$. One can see that the formula \eqref{higgs_tau_largedM} gives very good approximation (see also Fig. \ref{fig:nokinks_trans_single_tau}). At large $\Delta m$ we indeed have 
$\sqrt{2} \sigma  \approx m_0$.

\subsection{Quasivacua \label{sec:quasivacua_higgs}}

%
\begin{figure}[h]
    \centering
    \begin{subfigure}[t]{0.5\textwidth}
		\centering
		\includegraphics[width=\linewidth]{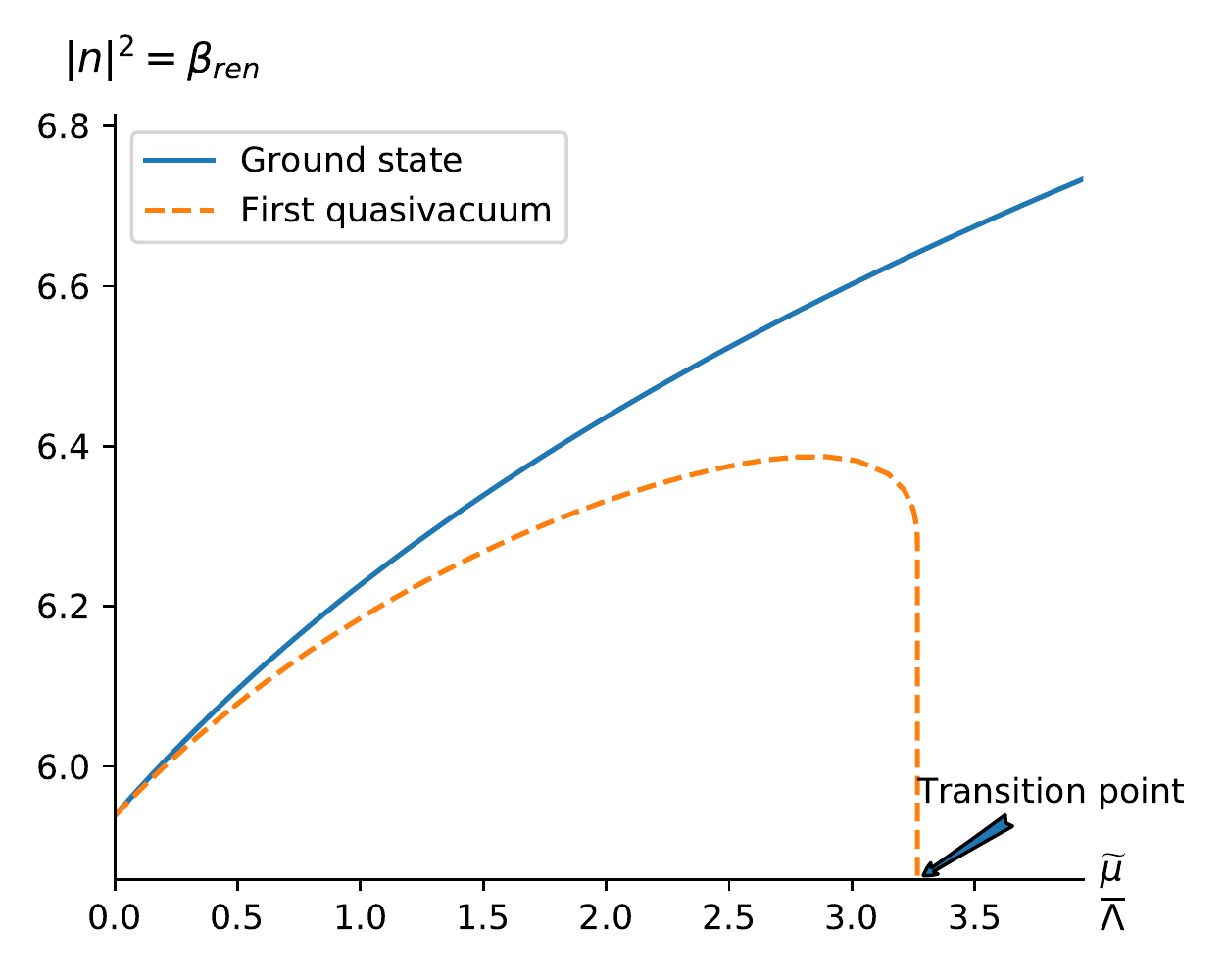}
		\caption{$|n|^2$}
	\label{fig:nokinks_trans_single_nsq} 
    \end{subfigure}%
    ~ 
    \begin{subfigure}[t]{0.5\textwidth}
		\centering
		\includegraphics[width=\linewidth]{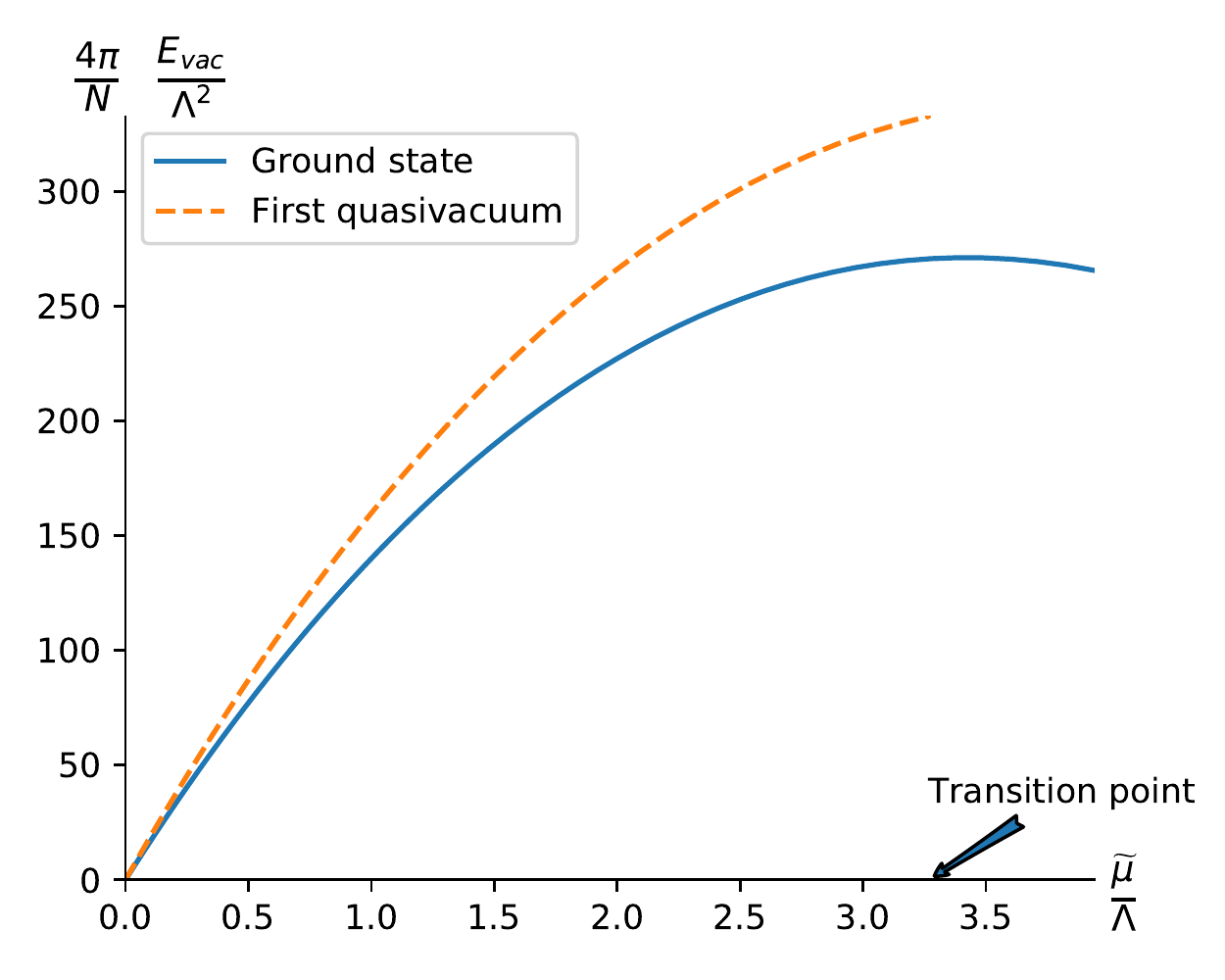}
		\caption{(Quasi)vacuum energy}
	\label{fig:nokinks_trans_single_Veff} 
    \end{subfigure}%
\caption{
	Example of Kinks-NoKinks phase transition for $\Delta m / \Lambda = 10$. Blue solid line refers to the true ground state $i_0 = 0$, orange dashed line represents the first quasivacuum $i_0 = 1$. Value of the deformation parameter $\wt{\mu}$ is on the horizontal axis (in the units of $\Lambda$), the phase transition point is indicated by an arrow.
	Both figures are the result of numerical calculations at $\Delta m / \Lambda = 10$, $N = 16$
}
\label{fig:nokinks_trans_single} 
\end{figure}
%
%
\begin{figure}[h]
    \centering
    \begin{subfigure}[t]{0.5\textwidth}
        \centering
        \includegraphics[width=\textwidth]{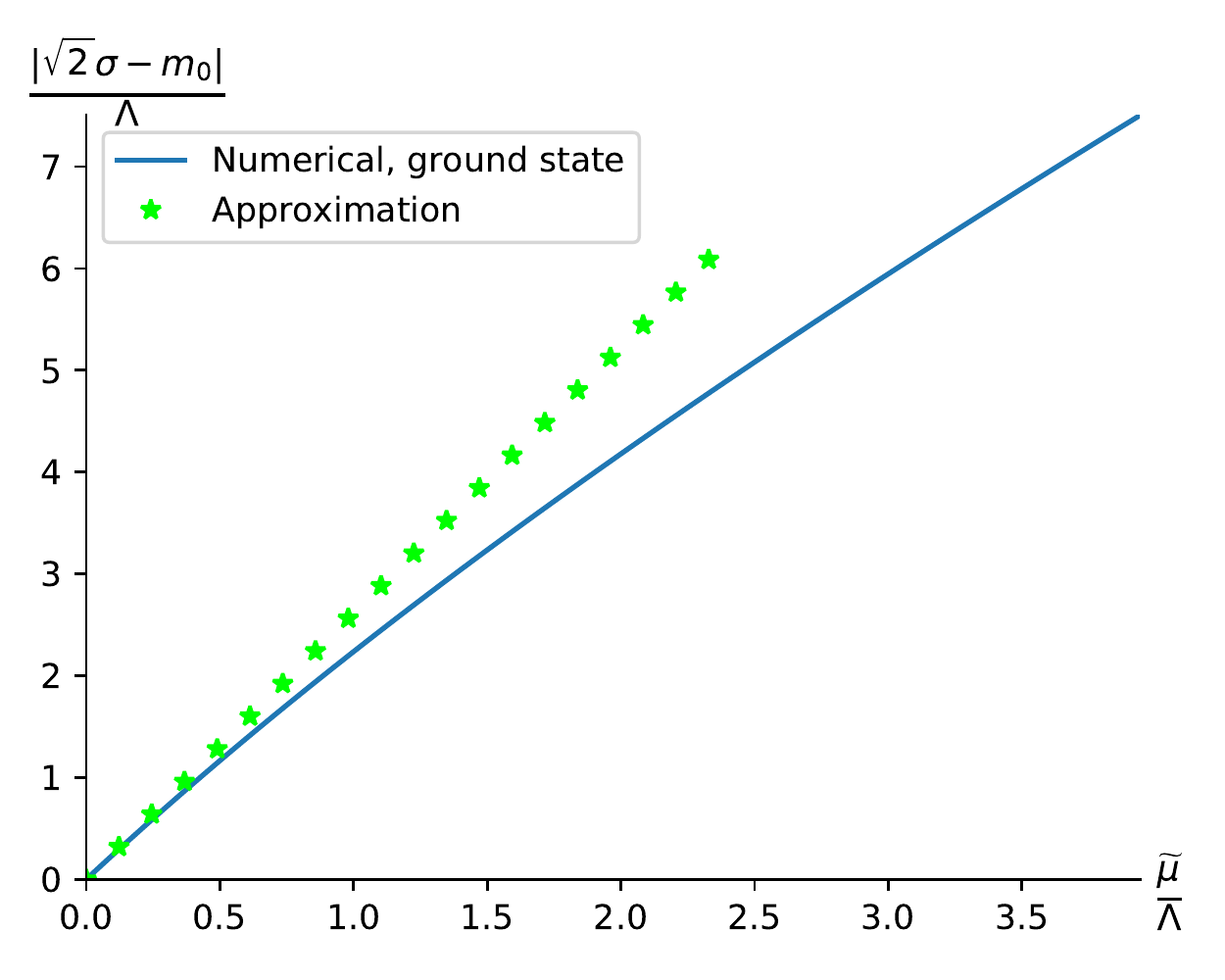}
        \caption{
        	$|\sqrt{2} \sigma - m_0|$ at small $\wt{\mu}$.
        }
    \label{fig:nokinks_trans_single_tau} 
    \end{subfigure}%
    ~ 
    \begin{subfigure}[t]{0.5\textwidth}
        \centering
        \includegraphics[width=\textwidth]{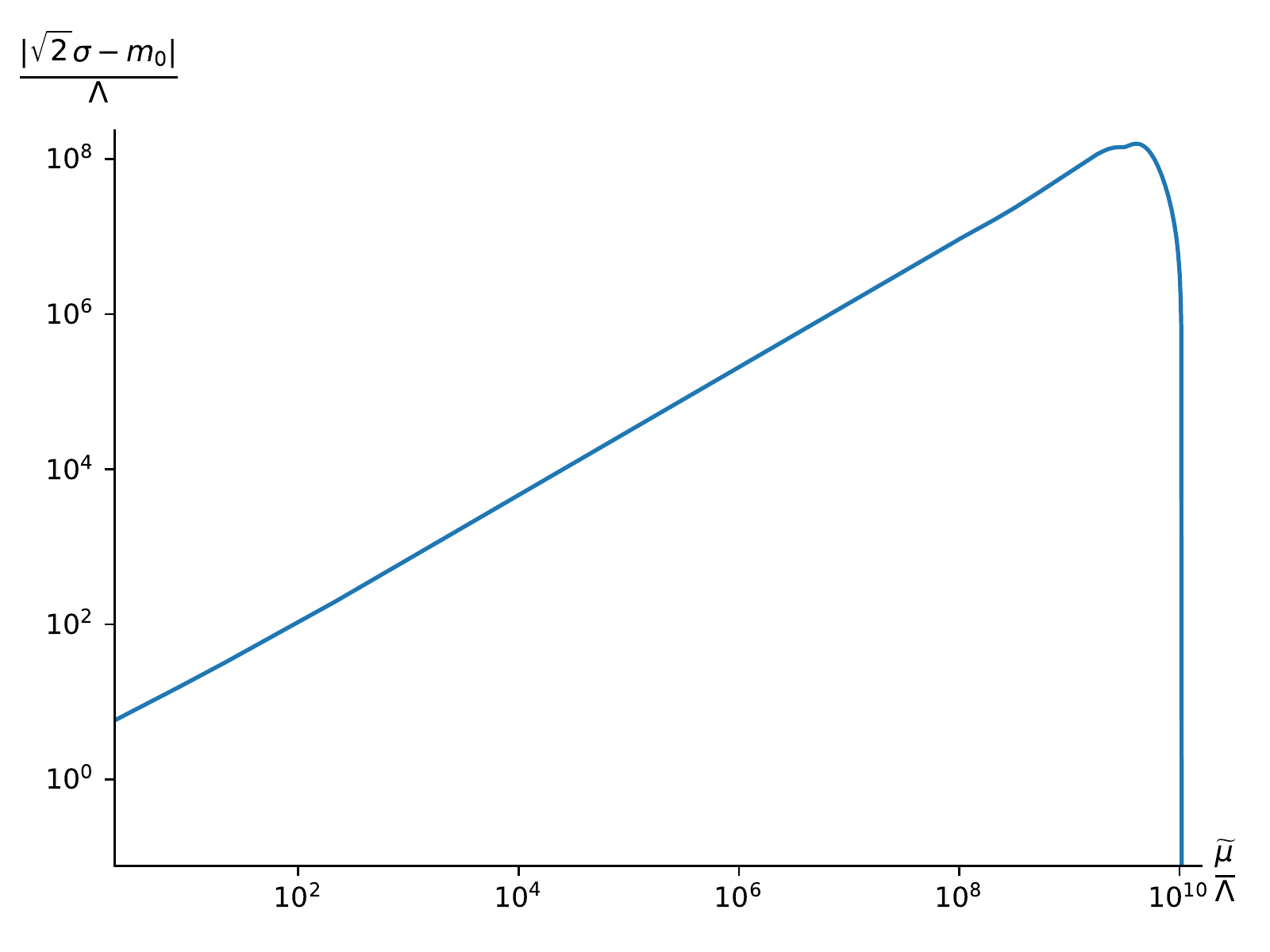}
        \caption{
        	$|\sqrt{2} \sigma - m_0|$ at large $\wt{\mu}$.
        }
    \label{fig:higgs_large_mu} 
    \end{subfigure}%
\caption{
	VEV of $\sqrt{2} \sigma - m_0$ at different scales.
	Figure \subref{fig:nokinks_trans_single_tau} shows small $\wt{\mu}$. Solid blue line is the result of numerical calculations, green stars show the approximate formula \eqref{higgs_tau_largedM}.
	Figure \subref{fig:higgs_large_mu} shows large-$\wt{\mu}$ behavior (in double log scale). One can see that as $\wt{\mu} \to m_G$ we indeed have $\sqrt{2} \langle \sigma \rangle \to m_0$.	
	The plots were made for fixed $\Delta m / \Lambda = 10$, $m_G / \Lambda = 10^{10}$, $N = 16$
}
\label{fig:higgs_tau_VEV} 
\end{figure}
\begin{figure}[h]
    \centering
    \begin{subfigure}[t]{0.5\textwidth}
        \centering
        \includegraphics[width=\textwidth]{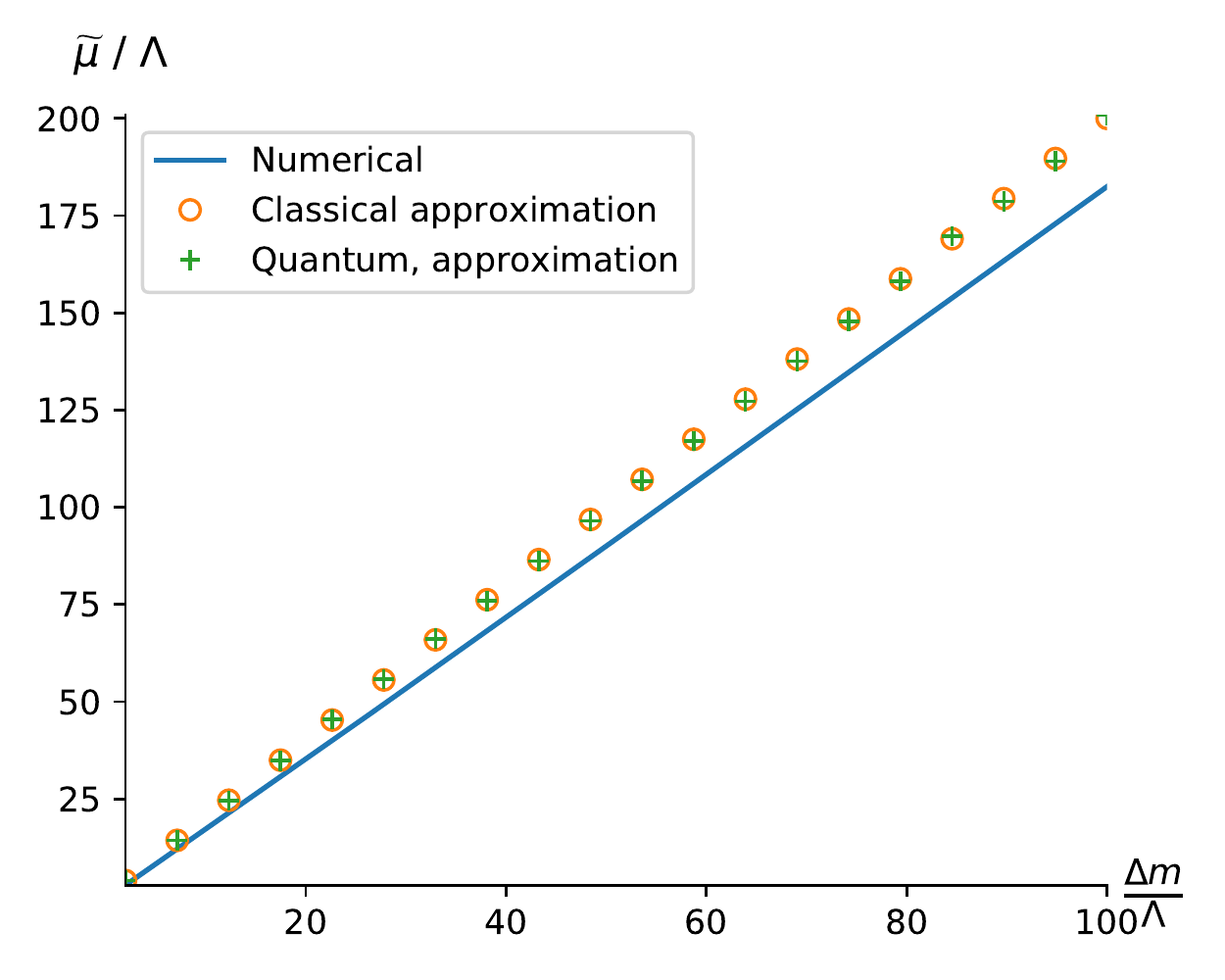}
        \caption{Agreement with the classical formula \eqref{nokink_trans_classic} if we set $\lambda = 0$}
        \label{fig:nokinks_trans_nolam}
    \end{subfigure}%
    ~ 
    \begin{subfigure}[t]{0.5\textwidth}
        \centering
        \includegraphics[width=\textwidth]{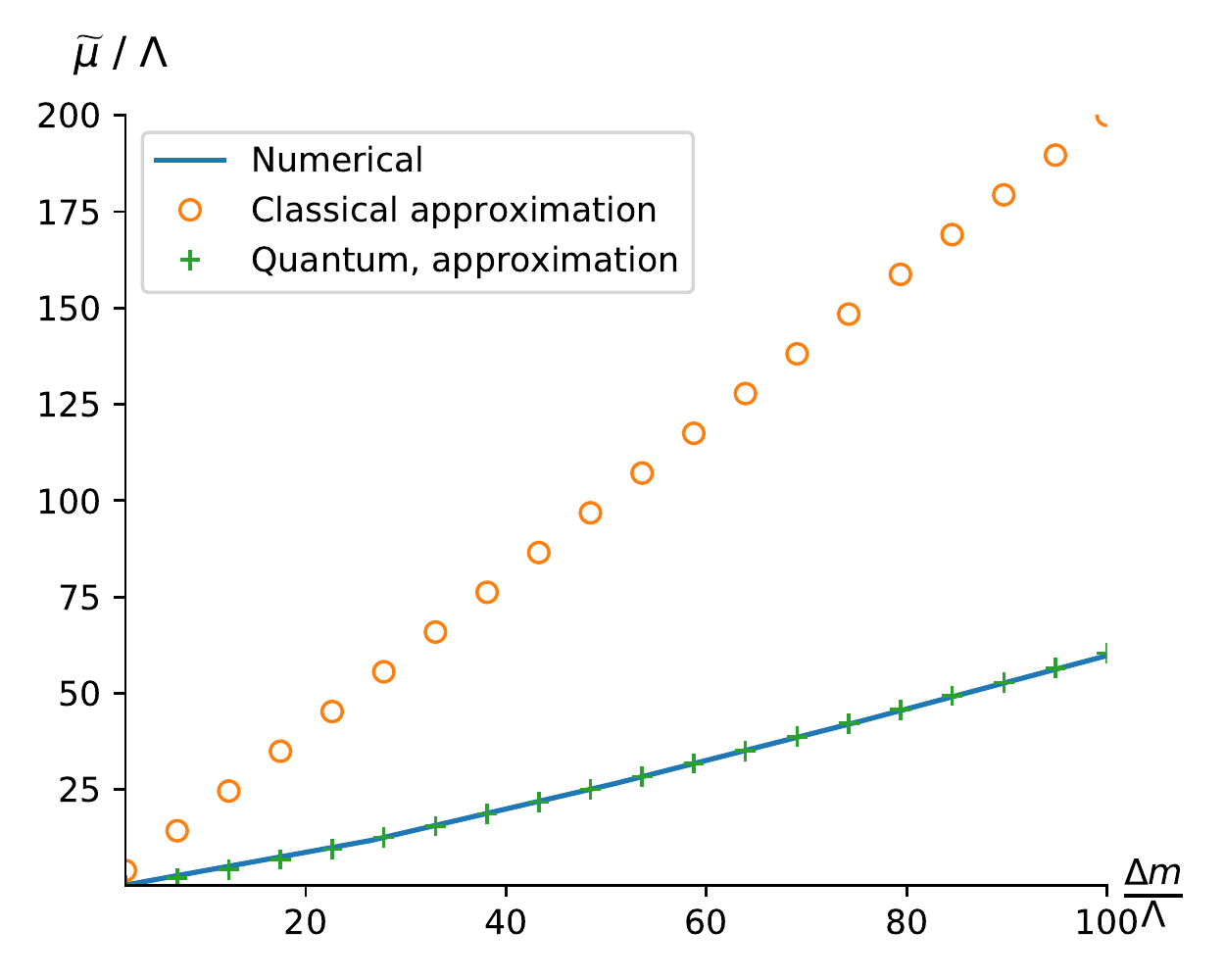}
        \caption{Presence of $\lambda$ magnifies the effect}
        \label{fig:nokinks_trans_lam}
    \end{subfigure}%
\caption{
	Kinks-NoKinks phase transition line. $\Delta m$ on the horizontal axis, $\wt{\mu}$ on the vertical axis. Solid blue line is the result of numerical calculation of the curve where all quasivacua have decayed leaving single true ground state.
	Orange circles represent the classical formula \eqref{nokink_trans_classic}, green \textquote{+} are the quantum approximation \eqref{nokink_trans_quant}.
	Figure \subref{fig:nokinks_trans_nolam} shows that if we set $\wt{\lambda}_0 = 0$ we indeed get good agreement with the classical formula \eqref{nokink_trans_classic}. However, the real scenario (figure \subref{fig:nokinks_trans_lam}) is better described by formula \eqref{nokink_trans_quant}
}
\label{fig:nokinks_trans} 
\end{figure}

Solution \eqref{higgs_classical} is just one of the possible vacuum states in the Higgs phase. In the supersymmetric case $\wt{\mu} = 0$ there are  $N$ degenerate vacua as dictated by Witten index. In \cite{Bolokhov:2010hv} it was shown that the theory at large $\Delta m$ is in the Higgs phase where different components of the $N$-plet $n_i$ develop a VEV. These vacua are characterized by
\begin{equation}
	\langle \sqrt{2} \sigma \rangle = m_{i_0}, \qquad \langle|n_{i_0}|^2\rangle = 2 \beta \,,\quad i_0 = 0 \,,\, \ldots \,, N-1
\end{equation}
Moreover, there are kinks interpolating between these vacua.

As we switch on the deformation parameter $\wt{\mu}$, these vacua split, and at small $\wt{\mu}$ we have one true ground state 
\eqref{higgs_classical} and $N-1$ quasivacua. Let us first consider this picture from the classical Lagrangian \eqref{lagrangian_init}. The classical  potential is
\begin{equation}
	{\mathcal V_\text{cl}}(n, \sigma, D) =  i\,D\left(\bar{n}_i n^i -2\beta \right)
		+  \sum_i\left|\sqrt 2\sigma-m_i\right|^2\, |n^i|^2	+ \upsilon (\mu) \sum_i \Re\Delta m_{i0} |n^i|^2
\label{classic_potential}
\end{equation}
Let us derive the mass spectrum in the vicinity of a vacuum $\sqrt{2}\sigma = m_{i_0}$ for some $i_0$. Then $n^i,\ i \neq i_0 $ are small, while
\begin{equation}
	n_{i_0} = \sqrt{2\beta} + \delta n_{i_0}
\end{equation}
From the $D$-term condition
\begin{equation}
	\delta n_{i_0} \approx - \frac{1}{2 \cdot 2\beta} \sum_{i \neq i_0} |n^i|^2
\end{equation}
and the potential \eqref{classic_potential} becomes
\begin{equation}
\begin{aligned}
	{\mathcal V_\text{cl}} &\approx 
		\sum_{i \neq i_0} |m_i - m_{i_0}|^2 \, |n^i|^2 
		+ \upsilon (\mu) \sum_{i \neq i_0} \Re(m_i - m_0) |n^i|^2
		- \upsilon (\mu) \Re(m_{i_0} - m_0) \sum_{i \neq i_0} |n^i|^2 
	\\
	&= \sum_{i \neq i_0} |n^i|^2 \left[ |m_i - m_{i_0}|^2 + \upsilon (\mu) \Re(m_i - m_{i_0}) \right]
\end{aligned}	
\label{classic_potential_masses}	
\end{equation}
so that the mass of the $n^i$ particle is
\begin{equation}
	M_i^2 = |m_i - m_{i_0}|^2 + \upsilon (\mu) \Re(m_i - m_{i_0})
\end{equation}
If $M_i^2$ were to turn negative for some $i$, this would signify that the vacuum under consideration is unstable. This  happens for all $i_0\neq 0$ if the deformation is large enough because there are always some  $i$ with $\Re(m_i - m_{i_0}) < 0$. 

To be more concrete, consider our choice of the masses \eqref{masses_ZN}. Then for the vacuum $i_0 = 0$ we have $\Re(m_i - m_{0}) > 0$ for all $i \neq 0$, and this vacuum is stable. However, the vacua $0 < i_0 < N/2$ can be shown to become unstable when the deformation parameter hits the critical value
\begin{equation}
	\upsilon (\mu_{\text{crit,} i_0}) 
		= 2 \Delta m\, \frac{1 - \cos(\frac{2 \pi}{N})}{\cos(\frac{2 \pi (i_0 - 1)}{N}) - \cos(\frac{2 \pi i_0}{N})}
		\approx \frac{4 \pi}{N} \, \frac{\Delta m}{\sin(\frac{2 \pi i_0}{N})} \, 
\label{nokink_trans_classic}		
\end{equation}
The last step is the large $N$ approximation. Similar statement holds for the quasivacua $N/2 < i_0 < N$, while the quasivacuum number $i_0 = N/2$ (for even $N$) 
decays when $\upsilon (\mu_{\text{crit,} N/2}) = 2 \, \Delta m$.
When $\wt{\mu}$ is above this critical value, the theory has unique vacuum, and there are no kinks left.

These quasivacua are seen from the one loop potential as well. Following \cite{Bolokhov:2010hv}, we can study these quasivacua as follows. Recall that deriving the effective potential \eqref{v_eff} we assumed that $n\equiv n_0$ can develop a VEV.
Now to study quasivacua we assume that $n_{i_0}$ is non-zero  and integrate out the other components of $n^i$.
Numerical calculation show that the resulting effective potential has a minimum for small deformations, but this minimum fades away at large $\wt{\mu}$, see Fig. \ref{fig:nokinks_trans_single}. On the plot \ref{fig:nokinks_trans_single_nsq}, this corresponds to the fact that $|n|^2$ rapidly drops near the phase transition point. Fig. \ref{fig:nokinks_trans_single_Veff} shows that the quasivacua are degenerate when supersymmetry is unbroken, and that the quasivacuum energy is indeed higher than that of the true ground state.

Figure \ref{fig:nokinks_trans} shows the phase transition curve. One can see that the classical formula \eqref{nokink_trans_classic} is valid only in if we set $\lambda = 0$ in \eqref{lagrangian_init}, but it is completely inadequate when the fermions gain extra mass. As we see from Fig. \ref{fig:nokinks_trans_lam} massive fermions magnify the effect. 

Let us derive better theoretical formula for the phase transition curve.
Consider, for example, first quasivacuum $i_0 = 1$. Then in the expression for $\beta_\text{ren}$ \eqref{master1} we will have $\Delta m_{i1} = m_i - m_1$ instead of $\Delta m_{i0}$. Then $\Re\Delta m_{01} < 0$, and for the phase transition point we can take roughly the point when $\beta_\text{ren} \to -\infty$, i.e.
\begin{equation}
	iD + \upsilon(\wt{\mu})\, \Re\Delta m_{01} + |\sqrt{2}\sigma - m_0|^2 = 0.
\end{equation}
Using \eqref{upsilon}, \eqref{lambda}, \eqref{higgs_tau_largedM} and an analog of \eqref{master2} one can show that the phase transition occurs at the point
\begin{equation}
	\wt{\mu}_\text{crit} \approx \frac{2 \, \Delta m}{1 + \wt{\lambda}_0 \frac{\ln {m_G / \Delta m}}{\ln{\Delta m / \Lambda}}} \,.
\label{nokink_trans_quant}
\end{equation}

At very large values of $\wt{\mu}$ all but one vacua have decayed, and the world sheet theory flows to the non-supersymmetric model. In this limit the VEV of $\sqrt{2} \sigma$ is again tends to $m_0$. Indeed, at large $\wt{\mu}$ we can solve the vacuum equations \eqref{master1} - \eqref{master3} approximately, and using the expression for $\Lambda$ \eqref{Lam_2d}, we find that
\begin{equation}
	\sqrt{2} \sigma - m_0 \sim \frac{\Delta m\, m_G^2}{\wt{\mu}^2} \, \ln\frac{\wt{\mu}}{\Delta m}\ln\frac{\wt{\mu}}{ m_G}
\end{equation}
which vanishes at large values of $\wt{\mu}$. This is supported by numerical calculations, see Fig. \ref{fig:higgs_large_mu}.

\subsection{Strong - Higgs phase transition \label{sec:trans_strong-higgs}}

%
\begin{figure}[h]
    \centering
    \begin{subfigure}[t]{0.5\textwidth}
		\centering
		\includegraphics[width=\linewidth]{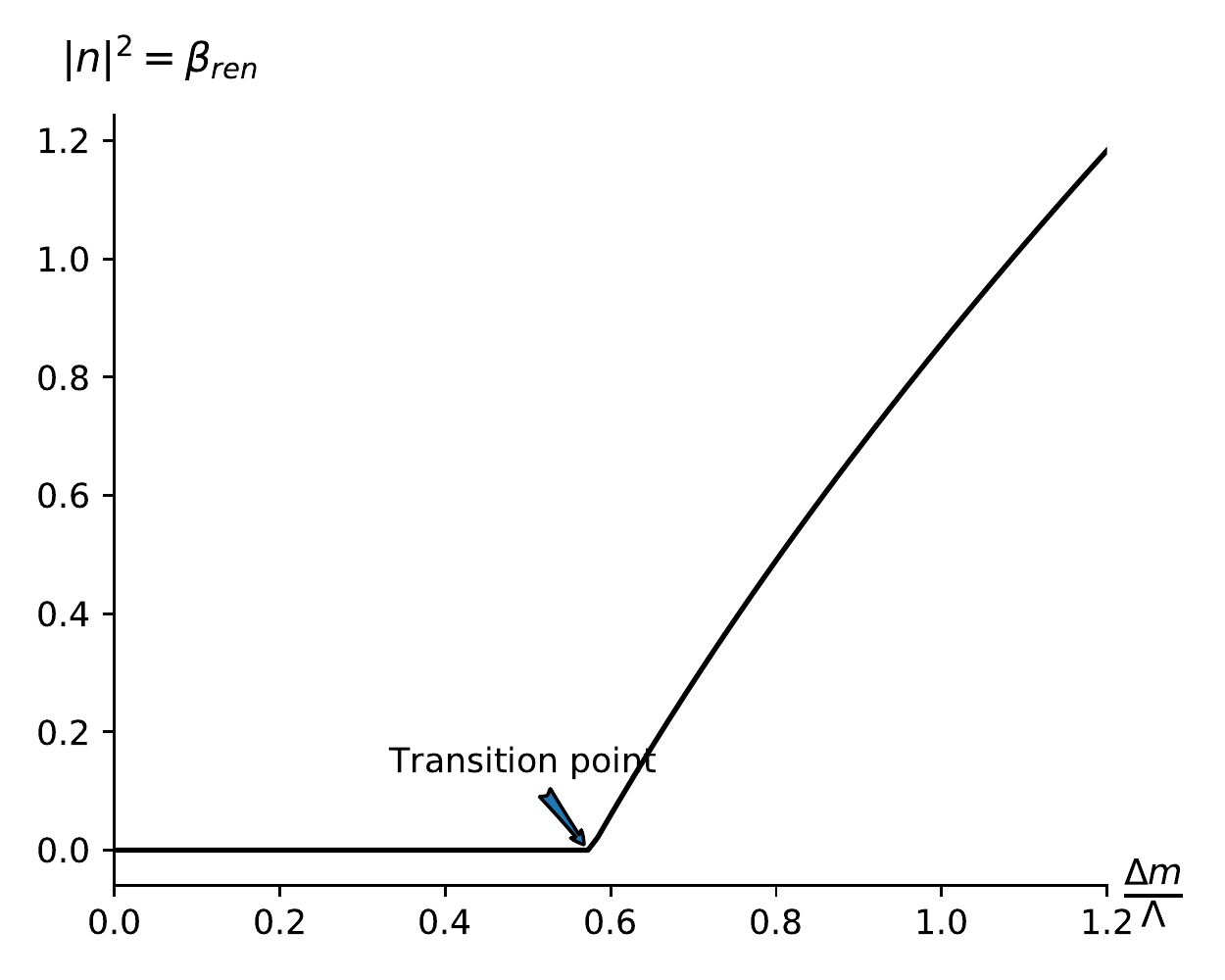}
		\caption{$|n|^2$}
	\label{fig:strong-higgs_single_nsq} 
    \end{subfigure}%
    ~ 
    \begin{subfigure}[t]{0.5\textwidth}
		\centering
		\includegraphics[width=\linewidth]{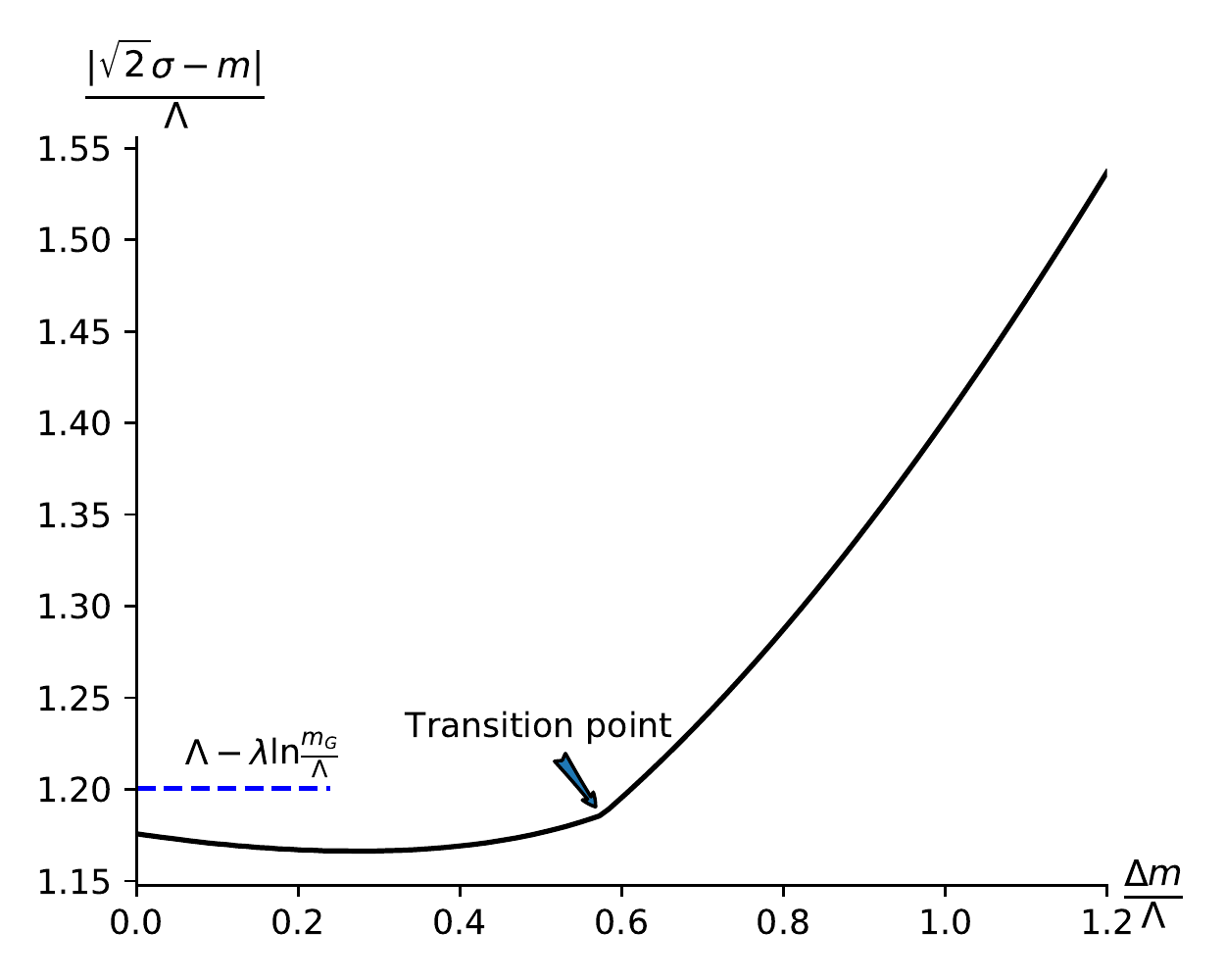}
		\caption{$\sqrt{2} \sigma - m$}
	\label{fig:strong-higgs_single_tau} 
    \end{subfigure}%
\caption{
	Strong - Higgs phase transition: VEVs. The curves show an example of the phase transition for fixed $\wt{\mu} / \Lambda = 0.03$, $N=16$. Mass scale $\Delta m$ is on the horizontal axis. Location of the phase transition point is indicated with an arrow. On the figure \subref{fig:strong-higgs_single_tau}, the position approximate strong coupling VEV \eqref{ground_state_correction}	is signified on the vertical axis by a blue dashed line.
	One can see that the character of the phase transition is qualitatively the same as in the pure non-supersymmetric \eqref{cpn_lagr_simplest} and supersymmetric \eqref{lagrangian_N=2} models, see \cite{Gorsky:2005ac,Bolokhov:2010hv}.
}
\label{fig:strong-higgs_single}
\end{figure}
\begin{figure}[h]
	\centering
	\includegraphics[width=0.7\linewidth]{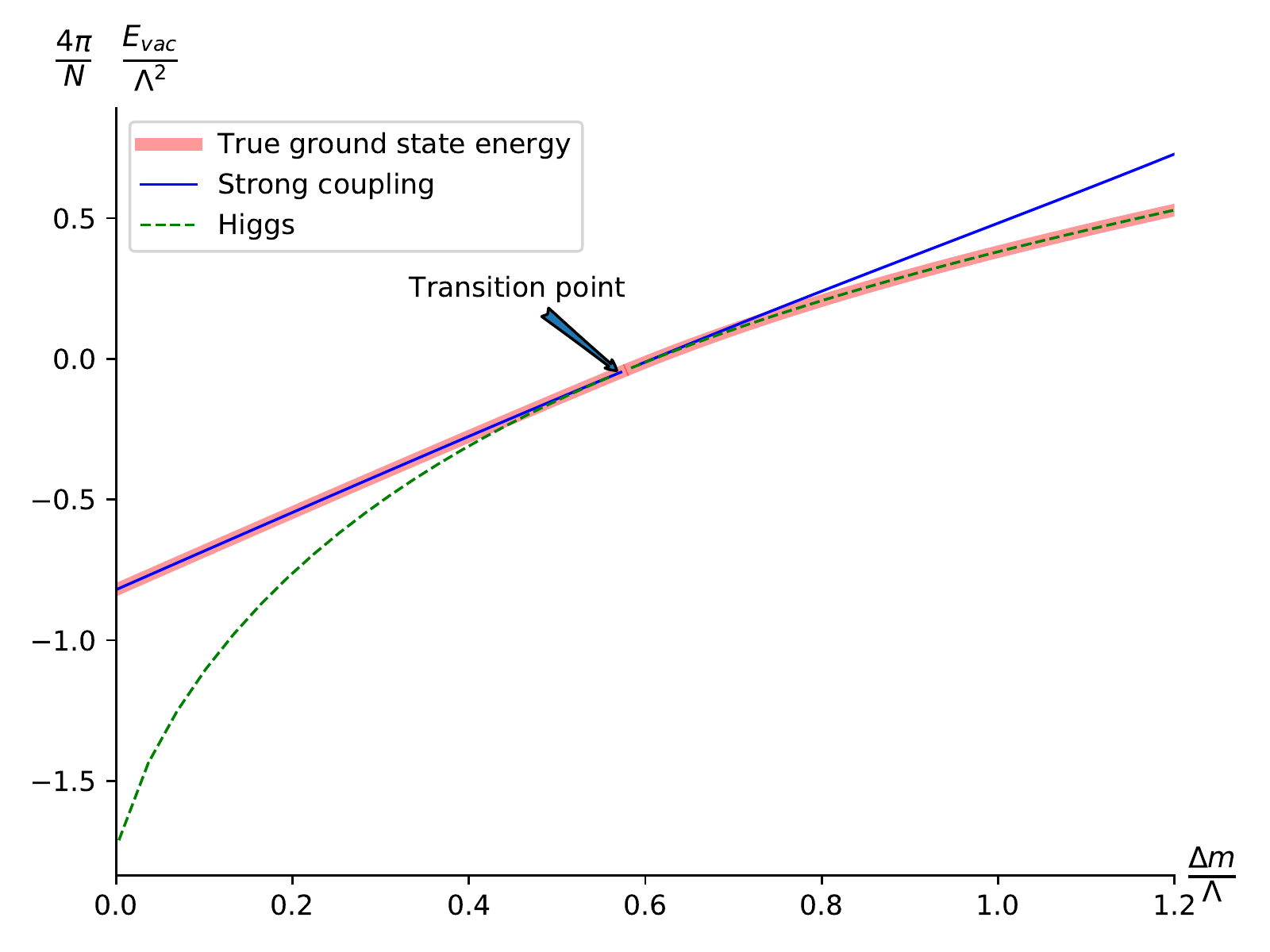}
	\caption{
		Strong-Higgs phase transition: energy. Red thick line is a numerical result for the ground state vacuum energy. Solid blue line to the right of the phase transition point is a numerical continuation of the strong coupling vacuum energy into the Higgs regime. Vice versa, dashed green line below the phase transition is a numerical continuation of the Higgs regime vacuum energy into the strong coupling (corresponds to the unphysical \textquote{state} with formally $|n|^2 < 0$. 
		At the phase transition point these two curves touch, and $|n|^2 = 0$.
		This plot is qualitatively the same as in the pure non-supersymmetric \eqref{cpn_lagr_simplest} and supersymmetric \eqref{lagrangian_N=2} models, see \cite{Gorsky:2005ac,Bolokhov:2010hv}
		In the numerical procedure we have set $\wt{\mu} / \Lambda = 0.03$, $N=16$
	}
\label{fig:strong-higgs_single_Veff} 
\end{figure}
\begin{figure}[h]
	\centering
	\includegraphics[width=0.7\linewidth]{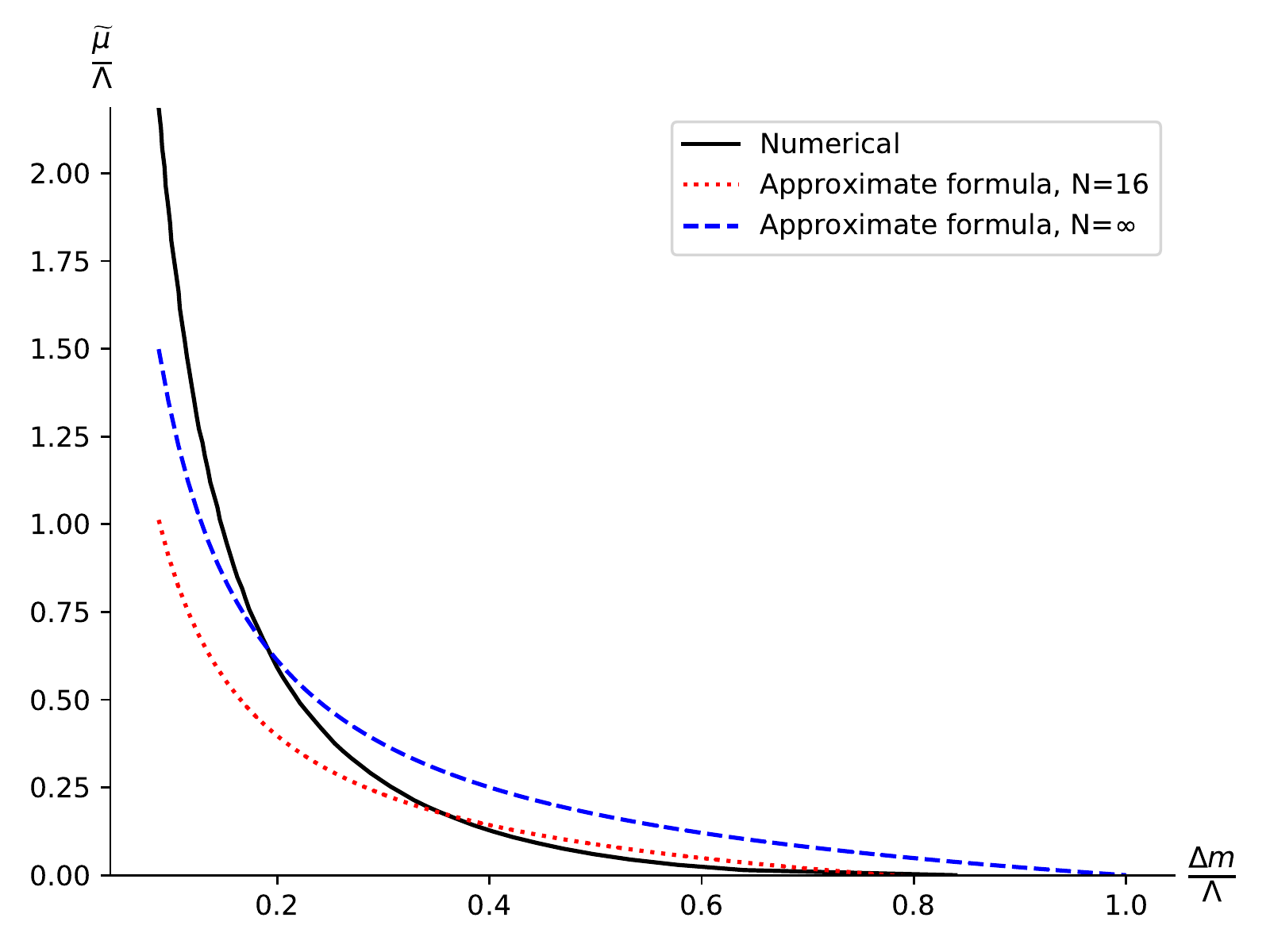}
	\caption{
		Strong-Higgs phase transition line. $\Delta m$ on the horizontal axis, $\wt{\mu}$ on the vertical axis.
		Solid black line is the numerical result for $N = 16$.
		Dotted red line is the $N=16$ approximate formula \eqref{strong_higgs_Nfin}.
		Dashed blue line is the $N\to\infty$ approximate formula \eqref{strong_higgs_Ninf}
	}
\label{fig:strong-higgs_trans} 
\end{figure}

It was found in \cite{Gorsky:2005ac, Bolokhov:2010hv} that 
for the non-supersymmetric \CP model \eqref{cpn_lagr_simplest}  and for the supersymmetric \CP model \eqref{lagrangian_N=2} a phase transition between strong coupling and Higgs phases occurs at the point $\Delta m = \Lambda$. At large $\Delta m$ the theory is weakly coupled and in the Higgs phase, while at small $\Delta m$ we have a strong coupling phase. 
We  expect similar behavior in our deformed model \eqref{lagrangian_init}.


Following  \cite{Gorsky:2005ac, Bolokhov:2010hv} we identify the Higgs-strong coupling  phase transition with a curve where $|n_0^2| =2\beta_\text{ren}$ turn negative. Thus we are looking for the solutions of the equation
\beq
\beta_\text{ren}=0,
\label{higgsphasetransit}
\eeq
where $\beta_\text{ren}$ is given by \eqref{master1}. 

In \ntwot supersymmetric model at $\wt{\mu} = 0$  a phase transition is at  $\Delta m = \Lambda$ \cite{Bolokhov:2010hv}.
The case $\wt{\mu} \neq 0$ is more complicated. We were not able to solve the vacuum equations \eqref{master1} - \eqref{master3} exactly, but an approximate calculation can be done in regions of small and very large $\wt{\mu}$.


First consider the region $\wt{\mu} \lesssim \Lambda$, and assume that the VEV of $\sigma$ is real-valued (this assumption is correct for the true ground state anyway). Then, using \eqref{master2} and the identity
\begin{equation}
	\prod_{k=1}^{N-1} \sin(\frac{\pi k}{N})  = \frac{N}{2^{N - 1}} \,,
\label{sinprod}	
\end{equation}
we can rewrite \eqref{master1} as
\begin{equation}
	2 \beta_\text{ren} = \frac{2 (N-1)}{4 \pi} \left( \ln\frac{\Delta m}{\Lambda} 
		+ \frac{1}{N-1} \ln N 
		+ \frac{1}{2} \ln\left(1 + \frac{\upsilon(\wt{\mu}) - 2(\sqrt{2}\sigma - m_0)}{2 \Delta m}\right) \right)
\end{equation}
Equating this to zero yields
\begin{equation}
	\upsilon(\wt{\mu}) - 2(\sqrt{2}\sigma - m_0) = 2 \Delta m \left( \left( \frac{\Lambda}{\Delta m} \right)^2 \, N^{- \frac{2}{N-1}} - 1 \right)
\label{trans_eq_Nfin}	
\end{equation}
At small deformations we can use the approximation $\upsilon(\wt{\mu}) \approx \wt{\mu}$, see \eqref{upsilon}. Moreover, in the strong coupling phase at fixed $\wt{\mu}$, the VEV of $\sigma$ does not depend on $\Delta m$ (this is exactly true in the supersymmetric and pure non-supersymmetric \CP models), and we can use $\Delta m = 0$ approximation \eqref{ground_state_correction} right up until the phase transition point. Then, from \eqref{trans_eq_Nfin} we can actually derive the equation for the phase transition curve:
\begin{equation}
	\wt{\mu}_\text{crit} = \frac{2 \displaystyle\frac{\Lambda^2}{\Delta m} \,  N^{- \frac{2}{N-1}} - \Lambda - \Delta m}{1 + 2 \, \wt{\lambda}_0 \ln\displaystyle\frac{m_G}{\Lambda}} \,,
\label{strong_higgs_Nfin}	
\end{equation}
or, sending $N \to \infty$,
\begin{equation}
	\wt{\mu}_\text{crit} = \frac{(2\Lambda + \Delta m) \, (\Lambda - \Delta m)}{\Delta m \left( 1 + 2 \, \wt{\lambda}_0 \ln\displaystyle\frac{m_G}{\Lambda} \right)}
\label{strong_higgs_Ninf}	
\end{equation}

These formulas give a good approximation for the phase transition curve, see Fig.~\ref{fig:strong-higgs_trans}. We see that with $\wt{\mu}_\text{crit}$ growing, $\Delta m_\text{crit}$ monotonically decreases. Moreover, comparing \eqref{strong_higgs_Nfin} and \eqref{strong_higgs_Ninf}, we can test the validity of our numerical calculations compared to the large-$N$ limit, as the numerics is done, of course, for a finite $N$\footnotemark.
\footnotetext{In the numerical calculations for this paper, we took $N$ = 16. Rough estimate of the accuracy from \eqref{strong_higgs_Nfin} and \eqref{strong_higgs_Ninf} is $1 - N^{- 1 / (N-1)} \approx 0.17$, i.e. qualitatively we can trust our results.}

In the region of large deformations, $\wt{\mu} \gg  m_G$.  We have
\begin{equation}
	2 \bren \sim \frac{N}{4\pi} \ln\frac{\upsilon (\mu_\text{crit})  \,\Delta m_\text{crit}  + \Delta m_\text{crit}^2 
	}{\Lambda_{2d}^2} =0,
\end{equation}
where $\Lambda_{2d}$ is exponentially small given by \eqref{beta_classical_largemu}.
From \eqref{upsilon} and \eqref{Lam_2d} we derive up to logarithmic factors
\beq
\Delta m_\text{crit} \sim \frac{(\Lambda_{4d}^{{\mathcal N}=1})^2 \ \wt{\mu}_\text{crit}}{m_G^2} \,
 \exp(- \text{const}\,\frac{\wt{\mu}_\text{crit}^2}{m_G^2}),
\label{strong-Higgs_critical}
\eeq
where  where we assumed that $\Delta m \ll m$, 
%
%
Here we see again that $\Delta m_\text{crit}$ monotonically decreases as $\wt{\mu}_\text{crit}$ becomes larger. 

\section{ Comments on 2d-4d correspondence}
\label{branes}

\subsection{Brane picture and 2d-4d matching conditions}

So far we have considered the  $\mu$-deformed 2d \CP model per se which is self-consistent. 
Let us now briefly comment on the requirements for the self-consistent treatment
of this 2d theory considered as the world sheet theory of the non-Abelian string in 
\ntwo supersymmetric QCD, deformed by a mass term $\mu$ for the adjoint matter. At $\mu=0$
the 2d-4d correspondence is seen from a different perspective. 
At the quantum level the situation is rather subtle
since the 4d instantons interfere with the 2d world sheet theory, which makes the problem 
quite nontrivial. 
Fortunately 
 it has been shown that  2d-4d correspondence  is seen  in the matching of RG flows of 2d and 4d  coupling
constants,  coincidence of 
spectra of BPS states \cite{dorey,SYmon,HT2},  and conformal dimensions at Argyres-Douglas  critical points \cite{tongAD}. 
A 
general discussion concerning the decoupling limits when the 4d degrees of freedom
do not influence the world sheet theory can be found in \cite{hori}.

In the brane engineering of \ntwo supersymmetric QCD the matching is nothing but the claim that one and the same 
object can not change if we look at
it
 as 4d or 2d observers. The standard IIA picture involves
two
 parallel NS5 branes, $N$ parallel D4 branes stretched between them and $N_f$ flavor 
branes which can be realized as semiinfinite D4' branes  or equivalently as KK monopoles \cite{witten}.
When lifted to the M-theory the whole configuration gets identified as  the single M5 brane wrapped around Seiberg-Witten
curve in the KK monopole background.
The non-Abelian string is represented by the D2 brane stretched between two NS5 branes along
some internal coordinate, say  $x_7$. Its length in this direction $\delta x_7$ 
coincides with 4d FI term  and yields the tension
of the dynamical non-Abelian string. The 4-4 strings yield the 4d gauge bosons, 4-4' strings
yield the chiral fundamental multiplets while 2-4' strings yield the hypers in 2d theory.
The 4d gauge coupling constant gets identified with the distance between NS5 branes 
$\delta x_6 \propto \frac{1}{g^2_{4d}}$ and exactly the same  geometric variable coincides
with the FI term in 2d theory. That is why the $\beta$ functions in 4d and 2d theories coincide
and geometrically reflect the back reaction of D4 and D4' branes on NS5 branes. The
matching of the spectrum of BPS states in 4d and 2d theories has the geometrical 
origin as well. The BPS dyons are represented as properly embedded into the 
brane geometry D2 branes which geometrically are seen as the dyonic kinks on the
world sheet of the non-Abelian string.

If we switch on the $\mu$ deformation one of the NS5 branes world sheet gets rotated.
Now the M5 brane is wrapped around holomorphic curve  $w= W'(v)$ in $(v,w,t)$ space
where $x_4+ix_5= v, \qquad x_8+ix_9 = w ,\qquad x_6 +i x_{10} =t$ and $W$ is
the
 superpotential of
the
 4d theory.
The embedding of  D2 or M2 brane representing the non-Abelian
string has been discussed in \cite{gukov}. It was argued that the M2 brane 
stretched between two M5 branes has a single possible stable embedding which fits with
the single vacuum in 2d theory.

In this paper we consider a bit different solution when the tension of the
string is fixed not by the FI term but by the mass of adjoint $T\propto \mu m$.
This means that there is no immediate identification of the string tension with
the distance between NS branes and therefore the  4d and 2d couplings can not
be immediately identified in the brane geometry as it was done in \ntwo case.
As expected from the brane picture the $\beta$ functions
in 4d and 2d theories indeed are not identical now and the non-perturbative scales are 
related in  a complicated manner \cite{YIevlevN=1}. It is better to use in this case
the IIB picture where the non-Abelian string is represented by a D3 brane with 
worldvolume coordinates $(x,t,D)$ where $D$ is the disc in the internal space
whose area yields the tension of the string.

One more input from the brane picture can be recognized if we  look more carefully at 
the background where the probe M2 brane is placed. Apart from the M5 boundary conditions
at the ends the background involves the $N_f$  KK monopoles in M theory.
Asymptotically the geometry involves the factors $C^2/Z_{N_F}$ due to the Taub-NUT metric induced by flavor branes.
Fortunately we can use some results concerning M2 probes in geometry involving Taub-NUT factors \cite{abjm}. 
The terms of interest are the Chern-Simons terms with opposite levels for $U(N)_k\times U(N)_{-k}$ for 
$N$ M2 branes. In our case we have $U(1)\times U(1)$ gauge group for  non-dynamical gauge field.
As was argued in \cite{abjm} a kind of  Higgs mechanism works and 
only the diagonal $U(1)$ survives. What is the remnant of the 3d
CS terms in 2d world sheet theory? We can define the  field $a(x)$ as
\beq
\exp(ia(x))= \exp \int A_z(x,z) dz
\eeq
where z is the  compact worldvolume coordinate of M2 brane.
The 3d CS term  induces the axion on the string world sheet
\beq
\delta L= N_f\int d^2 x a(x)*F
\eeq
Note that  CS term on M2 brane emerges as a result of the one-loop integration over the world sheet fermions 
which do not decouple completely, therefore
it can not be seen in the classical approximation. If we consider the representation of the non-Abelian
string via D3 brane in IIB the effective axion on the world sheet emerges as well in a similar manner. 
However the possible appearance
of the effective axion in 2d theory should be clarified accurately. We can not 
exclude that the CS and axion term disappears in the limit of complete decoupling 
of  4d degrees of freedom.

\subsection{Bulk criticality and world sheet theory}

One could question if the bulk criticality can be recognized in the
world sheet theory. Consider first large 
$N$
 limit  of \ntwo supersymmetric QCD
with $N=N_f$ and assume that all quark masses are equal. It was shown in \cite{zarembo}
making 
 use   of
the
  exact Seiberg-Witten solution that there is 
a 
 second order
phase transition at $m=2\Lambda$.  In particular 
it was shown that  there is 
a 
 jump of the 
derivative 
 of prepotential at this point.
We could question if
there are traces of this bulk phase transition in our string world sheet theory at $\Delta m=0$. In 
\ntwo theory the 4d-2d duality works well hence the phase transition 
in the world sheet could be  expected indeed. However, in the \ntwo limit of our U$(N)$ SQCD the average quark mass 
can be shifted away, therefore we cannot compare our results  with that of Ref.~\cite{zarembo}. The only hope is that we can 
identify a remnants of this phase transition at non-zero $\mu$.
The 4d phase transition cannot disappear if we switch on small parameter $\mu$,
however in order to  consider strings semiclassically we restrict ourselves to the weak coupling regime in 4d theory which
implies that parameter which fixes the string tension $\xi\propto \mu m$ is large.
 Therefore at small $\mu$ we are forced to assume very large $m$
and the phase transition point $m= 2\Lambda$ is far beyond of our approximation.
Nevertheless it would be very interesting to investigate the fate of phase transition 
found in \cite{zarembo} at large $\mu$.

One more example of bulk  criticality occurs if the average mass $m$ 
is very large but $\Delta m \propto \Lambda$. This is example of 4d Argyres-Douglas (AD) point
when the theory becomes superconformal. It was shown in \cite{tongAD} that  4d and 
2d AD points match perfectly and there is a matching of the critical
exponents as well. Our critical point at $\mu=0$ separating strong coupling and Higgs phases
matches  with a point $\Delta m= \Lambda$ on the curve (wall) of marginal stability. This curve form a circle 
$|\Delta m|= \Lambda$ in the large $N$ limit 
\cite{OlmezShifman}, and   2d AD point is a point on  this circle with  a non-zero phase.
If we switch on $\mu$ we expect that the  2d AD point
survives and at small $\mu$ evolves smoothly.
The clear-cut example of such smooth interpolation of AD point has been 
elaborated for $\mu$-deformed  $SU(2)$ gauge
group with $N_f=1$ in \cite{gvy} where it was identified as deconfinement
phase transition.

The final remark concerns the holographic picture which is possible since
we consider the large $N$ SQCD in the Veneziano limit. Instead of D3 brane 
in the background of NS5 and D5 branes in IIB supergravity  holographically we  consider D3
brane in peculiar 10d geometry with additional dilaton and form fields. 
The corresponding geometry has been identified
for pure \none limit at $\mu \rightarrow \infty$  in \cite{russoold,paredes} and in particular 
it reproduces the correct NSVZ beta function. For the generic $\mu$ the exact metrics is unknown 
but we expect  that the mass term amounts to the effective wall at the corresponding value of
the radial coordinate. 
A 
 very illuminating example
of
  how the bulk phase transition is seen 
on the probe string in the holographic framework has been found in \cite{russoholo}
for  ${\mathcal N}=2^*$ theory. In this case the exact holographic background is known
and coincides with the Pilch-Warner solution in IIB supergravity. The bulk theory enjoys the 
phase transition at strong coupling at mass of adjoint $M=\Lambda$. It turns out that
this phase transition indeed can be identified as criticality in the world sheet  theory of electric probe
string at the second order in the perturbation
theory in the inverse t'Hooft coupling at strong coupling. This clearly demonstrates 
that the 2d-4d matching of  bulk and world sheet
criticality is highly nontrivial in the holographic picture. It would be interesting
to investigate the non-Abelian magnetic string in the background  found in \cite{russoold,paredes}.

%
%

\section{Conclusions: Phase diagram  \label{Conclusions}}

%
\begin{figure}[h]
	\centering
	\includegraphics[width=1.0\linewidth]{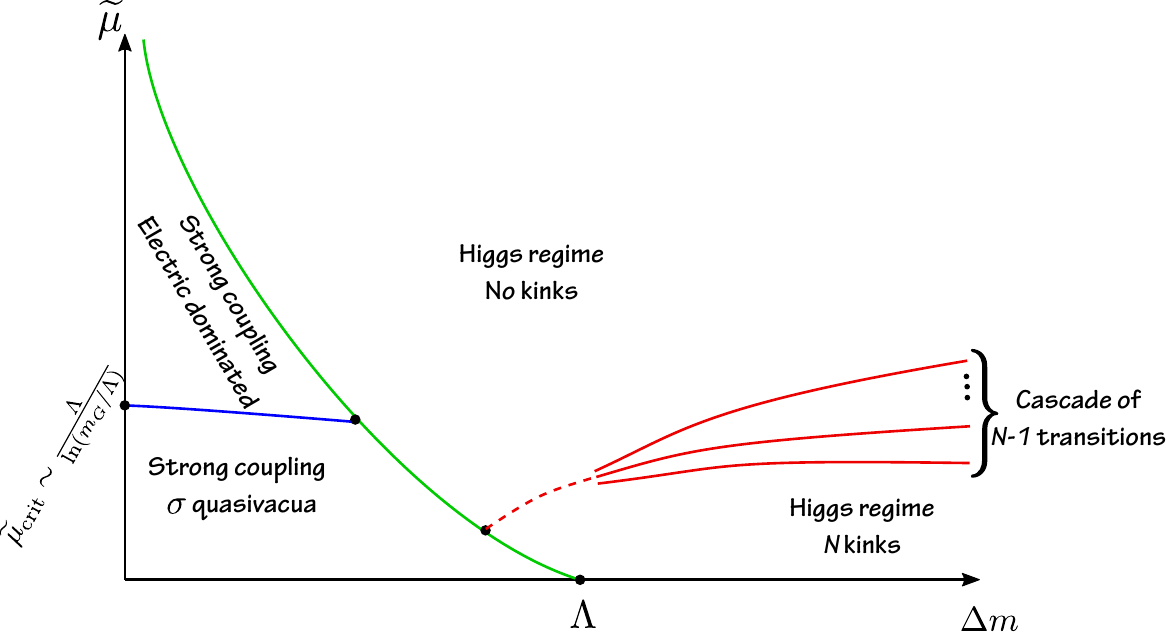}
	\caption{Whole phase diagram (schematically). $\Delta m$ on the horizontal axis, $\wt{\mu}$ on the vertical axis. Cascade of $N-1$ curves corresponds to the disappearance of kinks between the ground state and quasivacua. Dashed lines are drawn based on a general argument, since the $1/N$ expansion gives poor approximation in this region.}
\label{fig:allphases} 
\end{figure} 

In this paper we have studied dynamics of the $\mu$-deformed \CP model \eqref{lagrangian_init}. It arises as a world sheet theory of the non-Abelian string in \ntwo supersymmetric QCD, deformed by a mass term $\mu$ for the adjoint matter. When $\wt{\mu}$ is small, the two-dimensional theory is the \ntwot supersymmetric \CP model. As we increase the deformation parameter, the bulk theory flows to \none SQCD, while the world sheet theory becomes a non-supersymmetric $\mu$-deformed \CP model. This happens because fermion zero modes present in the bulk of the \ntwo theory are lifted when we switch on $\wt{\mu}$. As a consequence, at large $\wt{\mu}$ world sheet fermions become heavy and decouple, leaving us with the pure bosonic \CP model. In this paper we studied this transition in detail using the large $N$ approximation.

$\mu$-Deformed \CP model has two $N$-independent parameters, the deformation $\wt{\mu}$ (see \eqref{tildem}) and the mass scale $\Delta m$ which is the scale of the quark mass differences in the bulk theory. We obtained  a non-trivial phase diagram in the 
$(\Delta m, \,\wt{\mu} )$ plane, with two strong coupling phases  and  two Higgs phases separated by  three critical curves with two tricritical points. This phase diagram is shown on Fig. \ref{fig:allphases}. 

When $\wt{\mu}$ goes to zero, the supersymmetry is unbroken, and the theory is either in the strong coupling phase ( at small $\Delta m$ ) or in the Higgs phase (large $\Delta m$, weak coupling). In both phases there are $N$ degenerate vacua, and kinks interpolating between neighboring vacua are not confined. In the strong coupling phase at  small $\Delta m$ the photon becomes dynamical and 
acquire mass due to the chiral anomaly.

As we switch  on the deformation parameter   degenerate vacua split. 
At strong coupling we get a unique ground state and $N-1$ quasivacua, while the photon develops a small massless component. Kinks are now confined. When the deformation $\wt{\mu}$ is  small, the confinement is due to the splitting of the $\sigma$-quasivacua energies. As $\wt{\mu}$ gets larger eventually we cross the critical line where  original $\sigma$-quasivacua decay.  Now the quasivacua splitting and confinement of kinks is only due to the constant electric field.

In the Higgs phase at large $\Delta m$ the theory is at weak coupling. The $n$ field develop a VEV, photon is unphysical and heavy due to the Higgs mechanism.   When $\wt{\mu}$ is small enough
 energies of  $N$ degenerate vacua split, and  kinks interpolating between the neighboring quasivacua are confined. However as we increase $\wt{\mu}$ it crosses   critical  lines where (see e.g. \eqref{nokink_trans_classic})  quasivacua decay one by one leaving the theory with a single ground state, and thus without kinks. 

In this paper we have shown that results obtained in \cite{YIevlevN=1} for the  $\mu$-deformed  bulk theory  agree with the world sheet considerations. We can either go to the world sheet in the \ntwo theory and then take the large $\wt{\mu}$ limit, or first apply the large deformation in the bulk and then go to the world sheet theory.
In other words, the following diagram is commutative:
\begin{equation}
  \begin{tikzcd}[column sep=large]
  	\text{4d \ntwo SQCD} \arrow[swap]{r}{\text{worldsheet} } \arrow[swap]{d}{\text{large } \wt{\mu} } & \text{2d \ntwot \CP} \arrow[swap]{d}{\text{large } \wt{\mu} } \\
    \text{4d \none SQCD} \arrow[swap]{r}{\text{worldsheet} } & \text{2d } {\mathcal N}=0 \text{ \CP}      
  \end{tikzcd}	
\end{equation}
We note however that  a derivation of the world sheet theory at intermediate values of $\wt{\mu}$ is still absent.

As we already discussed we interpret kinks of the world sheet theory  as confined monopoles of the four-dimensional SQCD.
Our results show, in particular, that at large $\wt{\mu}$, when the bulk theory  basically becomes  \none SQCD,  monopoles survive  only in the strong coupling phase at very small mass differences below the critical line \eqref{strong-Higgs_critical}. In the Higgs phase quasivacua decay at large $\wt{\mu}$
which means that confined monopole and antimonopole forming a ''meson'' on the string ( see Fig.~\ref{fig:conf}) 
annihilate each other and disappear.
This confirms a similar  conclusion in \cite{YIevlevN=1}.

%
%

\section*{Acknowledgments}

The authors are grateful to M. Shifman for valuable discussions.
The work of A.G. was supported in part 
by Basis Foundation fellowship and RFBR grant 19-02-00214. 
A.G. thanks FTPI at University of Minnesota,  Simons Center for Geometry and Physics at Stony Brook
University and Kavli Institute at UCSB where the parts of work has been done 
for the hospitality and support.
The part of work of E. I. concerning the calculation of the diagrams and the strong coupling effective action was funded by Russian Foundation for Basic Research (RFBR) according to the research projects No. 18-32-00015 and No. 18-02-00048, and the rest of his work by the Foundation for the Advancement of Theoretical Physics and Mathematics \textquote{BASIS} according to the research project No. 19-1-5-106-1.
The work of A.Y. was  supported by William I. Fine Theoretical Physics Institute,   
University of Minnesota, Skolkovo Institute of Science and Technology and 
by Russian Foundation for Basic Research Grant No. 18-02-00048a.

%
%

\section*{Appendix A. Coefficients of the effective action}
\addcontentsline{toc}{section}{Appendix A. Coefficients of the effective action}
\renewcommand{\theequation}{A.\arabic{equation}}
\setcounter{equation}{0}

In this Appendix we give a brief overview of derivation of the effective action \eqref{effaction}.
Consider bosonic loops. In the Lagrangian \eqref{lagrangian_init} we can expand the $\sigma - n$ interaction term as
\begin{equation}
\begin{aligned}
	\left|\sqrt 2\sigma-m_i\right|^2 |n^i|^2 \approx 
		&\phantom{+} \left|\sqrt 2 \langle\sigma\rangle - m_i\right|^2  |n^i|^2	\\
		&+ 2 \Re(\sqrt{2}\delta\sigma) \cdot \left(\sqrt 2 \langle\sigma\rangle - \Re m_i\right)  |n^i|^2 \\
		&- 2 \Im(\sqrt{2}\delta\sigma) \cdot \Im m_i \,	|n^i|^2	
\end{aligned}		
\end{equation}
where $\delta\sigma$ are the vacuum fluctuations around the vacuum with $\Im\sigma = 0$. The diagram for the $\Re\sigma$ kinetic term is then proportional to 
$\left(\sqrt 2 \langle\sigma\rangle - \Re m_i\right)^2$, while the kinetic term for $\Im\sigma$ is proportional to $\left(\Im m_i\right)^2$. Calculation of the diagrams itself is straightforward.

Calculation of the fermion loops is a bit trickier. The fermion mass matrix can be read off from \eqref{lagrangian_init}. Say, for the flavor number $i$, 
\begin{equation}
	M_i = \left(\sqrt{2}\langle\sigma\rangle - \Re m_i \right) \cdot \text{Id} + i \left(\Im m_i\right) \cdot \gamma_\text{chir}
\label{fermion_mass}		
\end{equation}
where $\text{Id}$ is the $2 \times 2$ identity matrix, and $\gamma_\text{chir}$ is the two-dimensional analogue of the $\gamma_5$. This $\gamma_\text{chir}$ interferes with the traces over the spinorial indices. Say, the fermionic contribution to the $\Re\sigma$ kinetic term coming from the diagram on Fig.~\ref{fig:loops:scalar} is
\begin{equation}
\begin{aligned}
\begin{gathered}
	\includegraphics[width=0.15\textwidth]{sigma_sigma.pdf}
\end{gathered}	
	 &= - (i \sqrt{2})^2 \sum_i \int \frac{d^2 k}{(2\pi)^2} \Tr \left[ 
		\frac{i}{\slashed{k} - M_i} \frac{i}{\slashed{k} + \slashed{q} - M_i}
	\right] \\
	&= - 2 \sum_i  \int \frac{d^2 k}{(2\pi)^2} \Tr \left[ 
			\frac{\slashed{k} + M_i^\dagger}{k^2 - |M_i|^2} \frac{\slashed{k} + \slashed{q} + M_i^\dagger}{(k+q)^2 - |M_i|^2}
		\right]	\\
	&= - 4 \sum_i  \int \frac{d^2 k}{(2\pi)^2} \Tr \left[ 
			\frac{\left( k \cdot (k + q) \right) + \left(\sqrt{2}\langle\sigma\rangle - \Re m_i \right)^2 - \left(\Im m_i\right)^2}{(k^2 - |M_i|^2)((k+q)^2 - |M_i|^2)}
		\right]		
\end{aligned}	
\label{Re_sigma_ferm_init}	
\end{equation}
where $|M_i|^2 = \left(\sqrt{2}\langle\sigma\rangle - \Re m_i \right)^2 + \left(\Im m_i\right)^2$. Calculation of the integral itself is straightforward. The rest of the diagrams with fermionic loops are treated the same way. In the end we arrive at \eqref{eff_normalizations}.

Note that in the limit $\wt{\mu} \to 0$ supersymmetry is restored. In this case, in the vacuum $D = 0$, $\Im\sigma = 0$ we have
\begin{equation}
	M^2_{\xi_k} = m_{n_k}^2 = |\sqrt{2}\langle\sigma\rangle - m_k|^2
\end{equation}
and the coefficients \eqref{eff_normalizations} reduce to
\begin{equation}
	\frac{1}{e^2_{\Re\sigma}} = \frac{1}{e^2_{\Im\sigma}} = \frac{1}{e^2_{\gamma}} 
		= \frac{1}{4\pi}\,\sum_{k=0}^{N-1} \frac{1}{|\sqrt{2}\langle\sigma\rangle - m_k|^2}
\end{equation}
\begin{equation}
	b_{\gamma,\Im\sigma} = \frac{1}{2\pi}\,\sum_{k=0}^{N-1} \frac{1}{\sqrt{2}\langle\sigma\rangle - m_k}
\end{equation}

\section*{Appendix B. Photon mass}
\addcontentsline{toc}{section}{Appendix B. Photon mass}
\renewcommand{\theequation}{B.\arabic{equation}}
\setcounter{equation}{0}

In this Appendix we  diagonalize the photon-$\sigma$ mass matrix  in \eqref{effaction} to find the photon mass. In order to do that, let us write down bare propagators for $\Im\sigma$ and $A_\mu$ that follow immediately from \eqref{effaction} (in the Minkowski notation):
\begin{equation}
\begin{aligned}
	G^0_\gamma &= -i\, e^2_\gamma \, \frac{g^{\mu\nu} - \frac{k^\mu k^\nu}{k^2}}{k^2}
	\\
	G^0_{\Im\sigma} &= -\frac{i}{2}\, e^2_{\Im\sigma} \frac{1}{k^2 - \delta m_{\Im \sigma}^2}
\end{aligned}
\end{equation}
where we use the Landay gauge, while $\delta m_{\Im \sigma}^2$ is the contribution to the the mass of the $\Im\sigma$ field coming from the potential $V(\sigma)$ in \eqref{effaction}. In the vicinity of the true ground state \eqref{ground_state_correction} we have 
\begin{equation}
	\delta m_{\Im \sigma}^2 \approx 4 \lambda \Lambda \ln \frac{m_G}{\Lambda} \,.
\label{delta_Im_m_sigma}
\end{equation}
 At large $\wt{\mu}$, $\delta m_{\Im \sigma}^2 \sim  \lambda^2 \ln m_G / \Lambda$, see Sec. \ref{sec:strong_dm=0_largelam}.

\begin{figure}[h]
	\centering
	\includegraphics[width=0.6\linewidth]{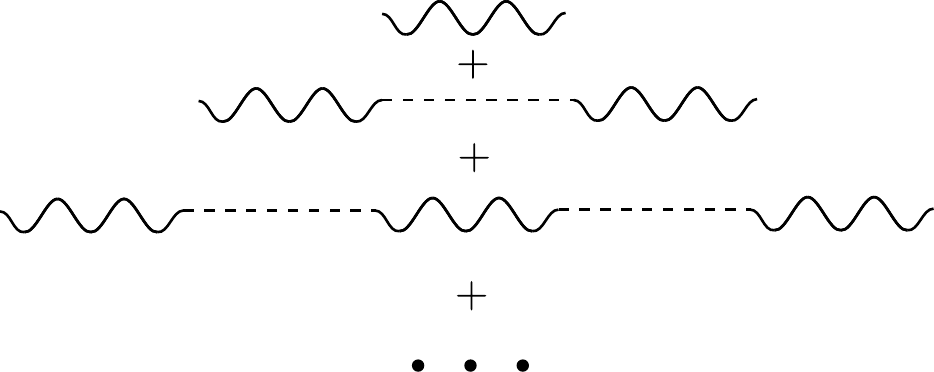}
	\caption{Contributions to the photon propagator}
\label{fig:iter_gamma} 
\end{figure}

Consider the photon propagator. Iterating the scalar $\Im\sigma$ insertions shown in Fig. \ref{fig:iter_gamma}, we obtain the  full photon propagator, 
\begin{equation}
\begin{aligned}
	\widehat{G}_\gamma 
		&= G^0_\gamma \frac{1}{1 - \frac{e^2_\gamma\, e^2_{\Im\sigma}\, b^2_{\gamma,\Im\sigma}}{k^2 - \delta m^2_{\Im\sigma}}}  \\
		&= -i\, e^2_\gamma \,  \left(g^{\mu\nu} - \frac{k^\mu k^\nu}{k^2}\right) \frac{k^2 - \delta m^2_{\Im\sigma}}{k^2\left(k^2 - \delta m^2_{\Im\sigma} - e^2_\gamma\, e^2_{\Im\sigma}\, b^2_{\gamma,\Im\sigma}\right)}  \\
		&= -i\, e^2_\gamma \,  \left(g^{\mu\nu} - \frac{k^\mu k^\nu}{k^2}\right) 
			\left(A\,\frac{1}{k^2} + (1 - A)\,\frac{1}{k^2 - \delta m^2_{\Im\sigma} - e^2_\gamma\, e^2_{\Im\sigma}\, b^2_{\gamma,\Im\sigma}}\right)
\end{aligned}
\label{photon_propagator_full}
\end{equation}
where the coefficient
\begin{equation}
	A = \frac{\delta m^2_{\Im\sigma}}{\delta m^2_{\Im\sigma} + e^2_\gamma\, e^2_{\Im\sigma}\, b^2_{\gamma,\Im\sigma}}
\end{equation}
increases from 0 to 1 as $\wt{\mu}$ runs from zero to infinity. 
What we see here is that at  non-zero $\wt{\mu}$, the photon acquires a  massless component. In the SUSY case (zero $\wt{\mu}$) the coefficient $A$ vanishes, and we have only the massive component. Note that the number of physical states do not change since
the massless photon has no physical degrees of freedom in two dimensions.
At large $\wt{\mu}$ the  massive component  becomes heavy and decouples ($A\to 1$). We are left with the massless photon much in the same way as in non-supersymmetric \CP model.

If we do a similar calculation for the $\Im\sigma$ propagator, we will get simply
\begin{equation}
\begin{aligned}
	\widehat{G}_{\Im\sigma} 
		&= G^0_{\Im\sigma} \frac{1}{1 - \frac{e^2_\gamma\, e^2_{\Im\sigma}\, b^2_{\gamma,\Im\sigma}}{k^2 - \delta m^2_{\Im\sigma}}}  \\
		&= -i\, e^2_{\Im\sigma} \, 	\frac{1}{k^2 - \delta m^2_{\Im\sigma} - e^2_\gamma\, e^2_{\Im\sigma}\, b^2_{\gamma,\Im\sigma}}
\end{aligned}
\label{Im_sigma_propagator_full}
\end{equation}
%
Just like in \cite{W79}, we see that the would-be massless phase of the $\sigma$ field acquires a mass
\begin{equation}
	m_{\text{arg}\, \tau}^2 =  \delta m^2_{\Im\sigma} + e^2_\gamma\, e^2_{\Im\sigma}\, b^2_{\gamma,\Im\sigma} \,.
\label{m_arg_sigma}
\end{equation}
This effect is taken into account by the additional term \eqref{DeltaV} in the effective potential \eqref{V_eff_strong_dm=0}.
At $\wt{\mu}=0$ $\delta m^2_{\Im\sigma}=0$ and the mass of the phase of $\sigma$  reduces to \eqref{wittenmass}.
Consider the leading correction at small $\lambda$.
For the ground state \eqref{ground_state_correction} at $\Delta m = 0$ we have
\begin{equation*}
	\frac{1}{e^2_{\Im\sigma}} \approx \frac{N}{4\pi\Lambda^2} \left(1 - 2\frac{\lambda}{\Lambda} \ln \frac{m_G}{\Lambda} \right) \,,
	\quad
	\frac{1}{e^2_{\gamma}} \approx \frac{N}{4\pi\Lambda^2} \left(1 - \frac{4}{3}\frac{\lambda}{\Lambda} \ln \frac{m_G}{\Lambda} \right) \,,
\end{equation*}
\begin{equation*}
	b_{\gamma,\Im\sigma} \approx - \frac{N}{2\pi\Lambda} \left(1 - \frac{\lambda}{\Lambda} \ln \frac{m_G}{\Lambda} \right) \,,
\end{equation*}
and, therefore,
\begin{equation}
	m_{\text{arg}\, \tau}^2 \approx 4 \Lambda^2 \left(  1 + \frac{7}{3} \frac{\lambda}{\Lambda} \ln \frac{m_G}{\Lambda} \right) \,.
\label{m_arg_sigma_leading}
\end{equation}

Let us look more closely at the photon propagator \eqref{photon_propagator_full} in the small $\wt{\mu}$ limit. We have  
\begin{equation}
	A \approx \frac{\lambda}{\Lambda}  \ln \frac{M}{\Lambda},
\end{equation}
and for the massless part of the photon propagator:
\begin{equation}
	\widehat{G}_{\gamma, \text{massless}} = 
		-i \,  \frac{g^{\mu\nu} - \frac{k^\mu k^\nu}{k^2}}{k^2} 
		\frac{4\pi}{N} \lambda\Lambda  \ln \frac{M}{\Lambda}
\end{equation}
From this Green function we calculate the electric field produced by a kink with electric charge $+1$ and  find for the vacuum energy splitting
\begin{equation}
	E_1-E_0 = \frac{1}{2 e_\gamma^2} F_{01}^2 = \frac{2\pi}{N} \left(\lambda \ln \frac{M}{\Lambda} \right)^2
\label{Evac_electical-1}	
\end{equation}
%

%
%


\clearpage


\end{document}